\documentclass[twocolumn,notitlepage,nofootinbib,floatfix,superscriptaddress]{revtex4-1}
\usepackage{amsmath, amsfonts, amsthm, amssymb, bbm}
\usepackage[margin=1in]{geometry}
\usepackage{algorithm,algpseudocode}
\usepackage{graphicx,xcolor,tikz}
\usepackage[colorlinks=true, hyperindex, breaklinks, linkcolor=blue, urlcolor=blue, citecolor=blue]{hyperref}
\usepackage{physics}
\usepackage{qcircuit}

\algblockdefx[ForEach]{ForEach}{EndForEach}[1]{\textbf{for each} #1 \textbf{do}}{\textbf{end for each}}

\graphicspath{{fewer_detections/figs/}}

\begin{document}

\preprint{APS/123-QED}

\title{High Rate Magic State Cultivation on the Surface Code}

\author{Yotam Vaknin}
\affiliation{AWS Center for Quantum Computing, Pasadena, CA 91125, USA}
\affiliation{Racah Institute of Physics, The Hebrew University of Jerusalem, Jerusalem 91904, Givat Ram, Israel}
\author{Shoham Jacoby}
\affiliation{AWS Center for Quantum Computing, Pasadena, CA 91125, USA}
\affiliation{Racah Institute of Physics, The Hebrew University of Jerusalem, Jerusalem 91904, Givat Ram, Israel}

 \author{Arne Grimsmo}
\affiliation{AWS Center for Quantum Computing, Pasadena, CA 91125, USA}
\author{Alex Retzker}
\affiliation{AWS Center for Quantum Computing, Pasadena, CA 91125, USA}
\affiliation{Racah Institute of Physics, The Hebrew University of Jerusalem, Jerusalem 91904, Givat Ram, Israel}

\date{\today}

\begin{abstract}

Magic state cultivation is a leading approach for generating the resource states required for fault-tolerant quantum computation. Here we present a new cultivation protocol that increases the success probability of magic-state generation in platforms with flexible and non-local connectivity. Our method implements cultivation directly on the surface code, avoiding the detour through alternative, less efficient error-correcting codes used in prior approaches. This both improves the acceptance rate and preserves compatibility with the geometry of the code and the hardware. Numerical simulations show that our protocols improve success probabilities and reduce output error rates compared with protocols tailored to locally connected platforms. Under realistic noise models for cold-atom and trapped-ion systems, we improve the rate of magic state generation by more than a factor of $20$. Finally, we study qubit loss and erasure, and show that very low error rates can be achieved with minimal overhead, reaching below $10^{-6}$ infidelity using only nine physical erasure qubits.
\end{abstract}
\maketitle

\section{Introduction}

Quantum error correction offers the potential to perform highly accurate quantum computations even when using noisy physical qubits \cite{aharonov1997fault,knill1998resilient}. Executing arbitrary quantum computations this way requires the generation of a state in a class of states called magic states. Generating a magic state requires unique machinery, different from standard error correction \cite{eastin2009restrictions, lao2022magic, gidney2023cleaner, bravyi2012magic, litinski2019magic}. Early estimates predicted that roughly $95\%$ of the physical qubits of a quantum computer would be designated for magic state generation \cite{Fowler_2012}, but many subsequent studies reduced these numbers considerably \cite{dennis2001toward, goto2014step, goto2016minimizing, chamberland2020very, itogawa2024even,hirano2024leveraging, gupta2024encoding}, up to the recent magic state cultivation (MSC) paper \cite{gidney2024magic}, which achieved magic state generation with resources comparable to a CX gate. Resource estimates based on MSC suggest that between $10\%-50\%$ of the qubits are required for magic state generation \cite{gidney2025factor, zhou2025resource}, depending on the computer's architecture and the specifics of the implementation.

The MSC result assumes the limitations that often occur for superconducting qubits: qubits are arranged in a square grid with local connectivity, and experience similar levels of noise while participating in gates and while idling. This local connectivity motivates the use of an error-correcting code called the color code \cite{bombin2006topological, kubica2015unfolding, gidney2312new}, which has a simple implementation of the $H$ gate using only single-qubit gates, but generally performs worse than the surface code \cite{kitaev2003fault, Fowler_2012}. Because one typically wants to use the surface code for a significant part of the computation, Ref.~\cite{gidney2024magic} expands the color code into a surface code in a complex procedure that requires many additional qubits and substantial post-selection.

Our work investigates how platforms with effective all-to-all connectivity, such as cold-atom arrays and trapped ions, can improve magic state cultivation. We also note that superconducting architectures could benefit from adding a limited number of long-range connections \cite{niu2023low, yoder2025tourgrossmodularquantum,marxer2023long,wang2025demonstration,bravyi2024high}. Recent work \cite{chen2025efficient} showed that such systems can reduce the cost of magic state cultivation, but at the price of using codes with non-planar connectivity, which are more difficult to implement in practice, mainly when qubit movement is required. In contrast, we use the surface code together with three-qubit gates in a way that respects the geometry of both the code and the physical qubits. Three-qubit gates have been demonstrated in both cold-atom arrays~\cite{levine2019parallel, evered2023high} and trapped-ion systems~\cite{lu2019global}. However, even without direct physical implementation, the added overhead of compiling 3-qubit gates into sequences of single- and two-qubit gates is small, because they appear sparsely throughout the protocol. We further observe that, when we adopt a realistic noise model for such devices, our scheme yields an additional order-of-magnitude reduction in the required resources. We therefore summarize below the specific contributions that enable this improvement.

We make three main contributions. First, we construct and simulate two magic-state cultivation protocols for the surface code based on transversal implementations of the gates $H$ and $H_{XY}$ using three-qubit gates. These schemes realize MSC-style magic state generation directly on the surface code, avoiding the color-code detour of Ref.~\cite{gidney2024magic} and yielding substantial savings in the magic-state generation rate on architectures with effective all-to-all connectivity. Second, we show that, when combined with a realistic noise model for cold-atom and trapped-ion devices, the $H$- and $H_{XY}$-based protocols achieve an additional order-of-magnitude reduction in magic-state resources compared with superconducting-style assumptions. Third, we revisit the logical $\mathrm{CX}$-eigenstate protocol of Dennis et al.~\cite{dennis2001toward,gupta2024encoding}, recast it as a $\mathrm{CX}$ cultivation scheme on the surface code. Our simulations show that although this protocol is not competitive with our $H$ and $H_{XY}$ constructions in our setting, it is code-agnostic and may be attractive for error correcting codes where transversal $\mathrm{CX}$ is particularly natural.

Lastly, we consider the effect of erasure and loss on magic state cultivation. Similar to \cite{jacoby2025magic}, we show that the final logical fidelity is limited only by the non-identifiable noise, i.e. the residual Pauli rate. We describe a minimal protocol that can utilize significant erasure bias, and achieve below $10^{-6}$ infidelity using only 9 physical erasure qubits. 

Our paper is organized as follows: We begin by outlining the theory of magic-state cultivation in detail. Next, we define our concrete protocol based on three qubit gate implementations. We then present numerical results under two noise models tailored to different quantum computing architectures. Finally, we show numerical simulations of erasure-based implementations of our scheme.

\section{Magic State Cultivation}
\begin{figure}
    \centering
    \includegraphics[width=0.42\textwidth,
    trim={25mm 25mm 25mm 25mm}]{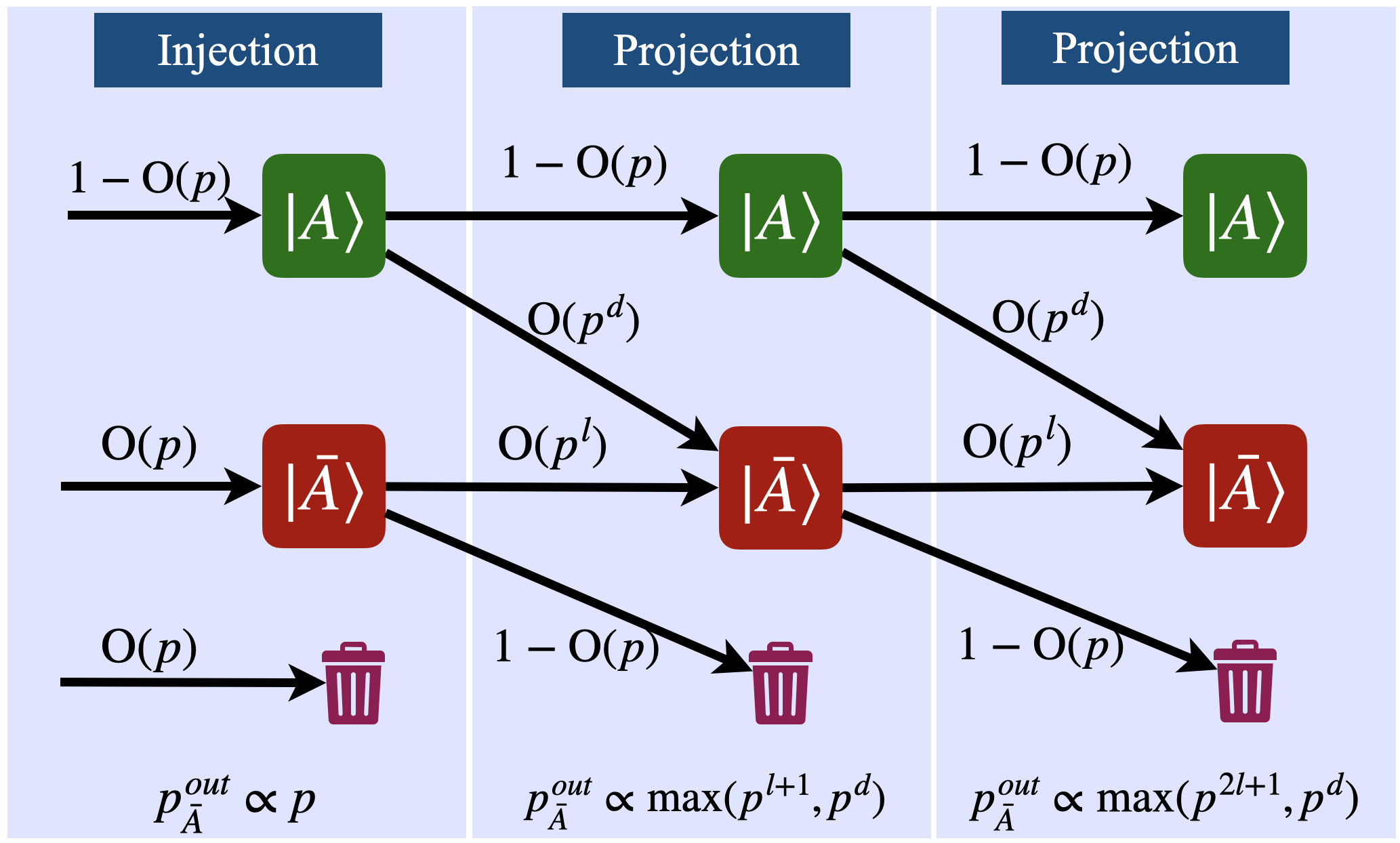}  
    \vspace{0.5cm}
\[
\Qcircuit @C=1em @R=.7em {
&\lstick{\ket{+}} & \ctrl{1} & \ctrl{2}& \ctrl{3} &\qw&\qw &\ctrl{2} & \ctrl{1}& \measureD{X} \\
&\lstick{\ket{0}} & \targ    & \qw     & \qw& \ctrl{3}& \qw&\qw    & \targ     & \measureD{Z}\\
&\lstick{\ket{0}} & \qw      & \targ   & \qw&\qw&\ctrl{3}  &\targ      & \qw   & \measureD{Z}\\
&\qw              & \qw      & \qw     &  \gate{A_1} &\qw&\qw&\qw      & \qw     & \qw\\
&\qw              & \qw      & \qw     & \qw& \gate{A_2}&\qw &\qw      & \qw     & \qw\\
&\qw              & \qw      & \qw     &  \qw&\qw& \gate{A_3}&\qw      & \qw     & \qw
\inputgroupv{4}{6}{.8em}{1.9em}{\ket{\psi}}\\
}
\]
        \caption{{\textbf{Schematics of magic state cultivation} -- }
    \label{fig:phase_kickback}
    (Top) Partial Markov chain description of the cultivation protocol. Arrows are labeled by the leading-order scaling of their transition probabilities. We write $p_{\bar{A}\to{A}}\propto p^l$, where $l=1$ for phase-kickback and $l=2$ for \textit{double-checking}. The final fidelity is limited by $p^d$, the same scaling as the error detecting distance of the underlying code. (Bottom) Schematic circuit for a single phase-kickback projection using a GHZ ancilla and a transversal implementation of $A$ on a quantum error correcting code. The $X$-basis measurement of the top ancilla qubit measures the eigenvalue of $A$, while the $Z$-basis measurements check the stabilizers of the GHZ state. We only keep runs in which all of these  measurements, together with a subsequent round of surface code stabilizer measurements, return the trivial outcomes.}
\end{figure}
Magic state cultivation builds on the observation that measurements can achieve extremely high fidelity by being repeated and by coupling the data qubit to multiple ancilla qubits \cite{shor1996fault}. Magic state preparation is inherently noisy: each attempt typically produces a state with a component outside the target magic state. The basic idea is to generate such a noisy magic state and repeatedly measure an observable $A$ for which the ideal magic state is a \(+1\) eigenstate. Conditioning on the \(+1\) outcome progressively increases our confidence that the state is correct. Ultimately, the fidelity will be constrained by the probability of flipping the state during measurement, which scales as the gate fidelity to the power of the code distance.   

Let \(A\) be a unitary and Hermitian single-qubit operator with eigenvalues \(\pm 1\). The \(+1\) eigenstate is the target magic state \(\ket{A}\), and \(\ket{\bar{A}}\) denotes the \(-1\) eigenstate. 

The process begins with a superposition or mixture of states $\ket{A}$ and $\ket{\bar{A}}$ and aims to increase the population of $\ket{A}$. This is achieved by repeatedly measuring the observable $A$. The process remains effective even when the measurement error rate is high, provided that measurement faults rarely change the underlying state without also producing an observable syndrome. Cultivation achieves this by supplementing the measurement with ancillary checks and a stabilizer measurement of the surface code. We abort and restart the protocol whenever any one of these checks reports a syndrome.

Even if we observe the correct measurement result with no syndromes, a noisy measurement of $A$ can still fail in two ways: (1) with probability $ p_{A\to \bar{A}}$, we can flip a correct $\ket{A}$ while measuring it, or (2) with probability $p_{\bar{A}\to \bar{A}}$ a wrong state $\ket{\bar{A}}$ is missed due to measurement errors.
This idea is illustrated in Fig.~\ref{fig:phase_kickback}.

The core of cultivation is that if there is a separation of scale between the two probabilities:
$$p_{\bar{A}\to \bar{A}} \gg p_{A\to \bar{A}},$$
repeated measurements improve the accuracy, up to a floor of the flipping probability $p_{A\to \bar{A}}$. 

The more precise result depends on the input probabilities and obeys:
$$
    p_{\overline{A}}^{out} \le  p_{A}^{in}\frac{p_{A\to\overline{A}}}{1-p_{\overline{A}\to\overline{A}}}+p_{\overline{A}}^{in}\left(p_{\overline{A}\to\overline{A}}\right)^{k} + \mathcal{O}(p^2_{A\to\overline{A}}). 
$$
Here $p_{\bar{A}}^{out}$ is the probability of incorrectly accepting a $\ket{\bar{A}}$ state, $p_{A(\bar{A})}^{in}$ is the probability of correctly starting the procedure in $\ket{A(\bar{A})}$ (i.e. successful injection), and $k$ is the number of repetitions. 

We define the probability of accepting a $\ket{A}$ state as $p_{A}^{out}$. The final state infidelity is defined as:
$$
\mathsf{IF} = \frac{p_{A}^{out}}{p_{A}^{out}+p_{\bar{A}}^{out}},
$$
and the acceptance rate as:

$$
\mathsf{Rate} =\left( p_{A}^{out}+p_{\bar{A}}^{out} \right).
$$
In our plots, we show the expected number of attempts per accepted shot, which is the inverse of the rate. In the next section, we show that for code with distance $d$, cultivation can achieve $\mathsf{IF}=\mathcal{O}(p^d)$ and $\mathsf{Rate}= 1- \mathcal{O}(p)$.

\subsection{Implementation on an error correcting code}

\begin{figure} 
    \centering
    \includegraphics[width=0.48\textwidth]{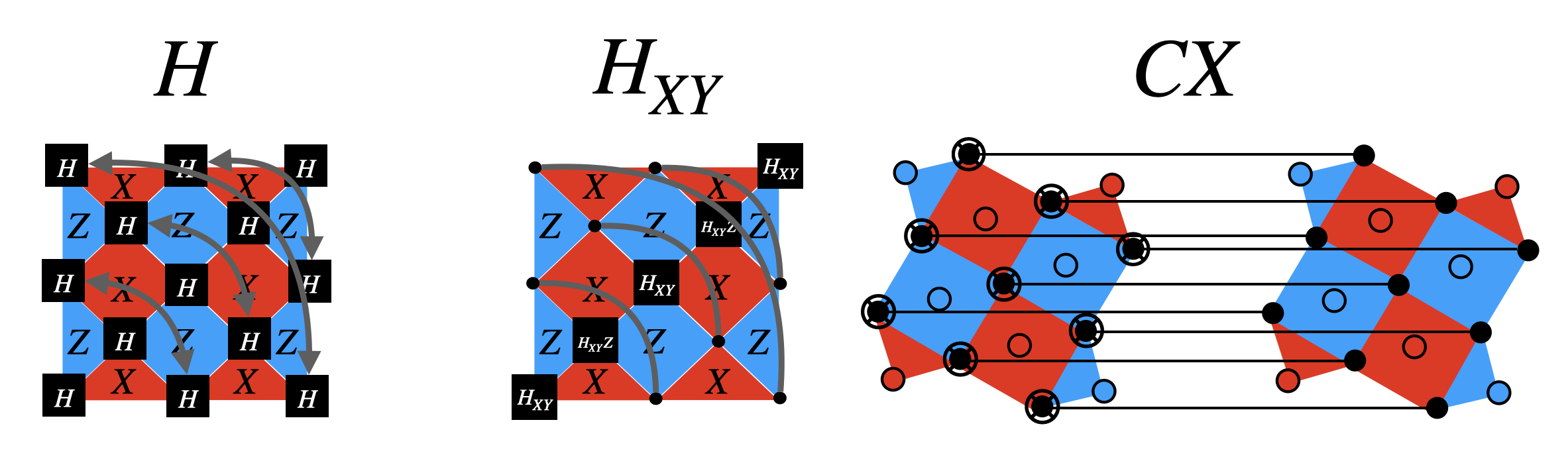}  
    
    \caption{{\textbf{Transversal implementation of $H$,$H_{XY}$ and $CX$ on the surface code} -- }
    \label{fig:transversal_operations}
     On the unrotated surface code, $H$ is implemented via a pattern of SWAP and single-qubit $H$ operations, while $H_{XY}$ is implemented using $H_{XY}$, $H_{XY}Z$, and $CZ$ gates. Logical $CX$ is transversal for both rotated and unrotated surface codes by applying physical $CX$ gates between corresponding data qubits of two code patches. The figure depicts the construction on rotated surface-code patches, but the transversal pairing applies to either geometry. }
\end{figure}

Consider a logical state $\ket{\psi_L}$ encoded in an error correcting code, and suppose the logical operator $A$ admits a decomposition into single- and two-qubit gates $\{A_i\}_i$, so that
\[
    A = \prod_i A_i.
\]
Note that $A$ acts as an automorphism of the stabilizer group. The existence of such a decomposition means that we can implement $A$ logically by acting only on individual physical qubits or pairs of qubits. We refer to such an $A$ as a transversal logical operator. 

Initializing an error correcting code in the magic state $\ket{A}$ requires a non-transversal unitary, which by the Eastin--Knill theorem~\cite{eastin2009restrictions} cannot be made fully fault tolerant. We therefore use a circuit that isn't fault-tolerant, called a state-injection circuit \cite{lao2022magic,gidney2023cleaner, jacoby2025magic}. State injection can produce an incorrect state with a probability that is linear in the physical error rate $p$.

To measure the $A$ operator, we couple the encoded state to a GHZ ancilla state through controlled-$A_i$ gates and then measure the GHZ state in the $X$ basis, as illustrated in Fig.~\ref{fig:phase_kickback}. We call this circuit  a phase-kickback measurement circuit. For phase-kickback, the logical error probabilities scale as:
\begin{equation}
\label{eq:logical_scaling_phase_kickback}
    p_{A \to \bar{A}} \propto p^{d}, \qquad
    p_{\bar{A} \to \bar{A}} \propto p,
\end{equation}
where $d$ is the distance of the underlying code. Any pattern of faults that doesn't produce a syndrome and maps the logical state from $\ket{A}$ to $\ket{\bar{A}}$ must involve at least $d$ physical errors, while incorrectly accepting an incoming $\ket{\bar{A}}$ state can be caused by a single fault in the measurement protocol.
Note that Eq.~(1) omits the probabilities of identifiable error events, which are discarded rather than accepted. Consequently, for input \(\ket{\bar A}\) the most likely outcome is post-selection, while for input \(\ket{A}\) it is acceptance; see Fig.~\ref{fig:phase_kickback}.
In practice, our best implementation uses a different circuit called \emph{double-checking}, defined in Section~\ref{sec:phase-kickback-and-double-checking}. It preserves the favorable scaling $p_{A \to \bar{A}} \propto p^{d}$ while improving the acceptance of incorrect states to:
\begin{equation}
\label{eq:logical_scaling_double_checking}
    p_{\bar{A} \to \bar{A}} \propto p^{2}.
\end{equation}
We establish these scalings by performing a complete search in error space, or when this becomes too computationally expensive, we run a full state-vector simulations and observe the same asymptotic behavior, see Appendix~\ref{SI:full_vector_simulation}.

These scalings imply that injecting a magic state into a distance-$3$ code achieves $p^d$ logical error with a single \textit{double-checking} round, and a distance-$5$ code achieves the same $p^d$ scaling with only two iterations of \textit{double-checking}.

\subsection{Transversal implementation on the surface code}
\begin{figure} 
    \centering
    \includegraphics[width=0.45\textwidth]{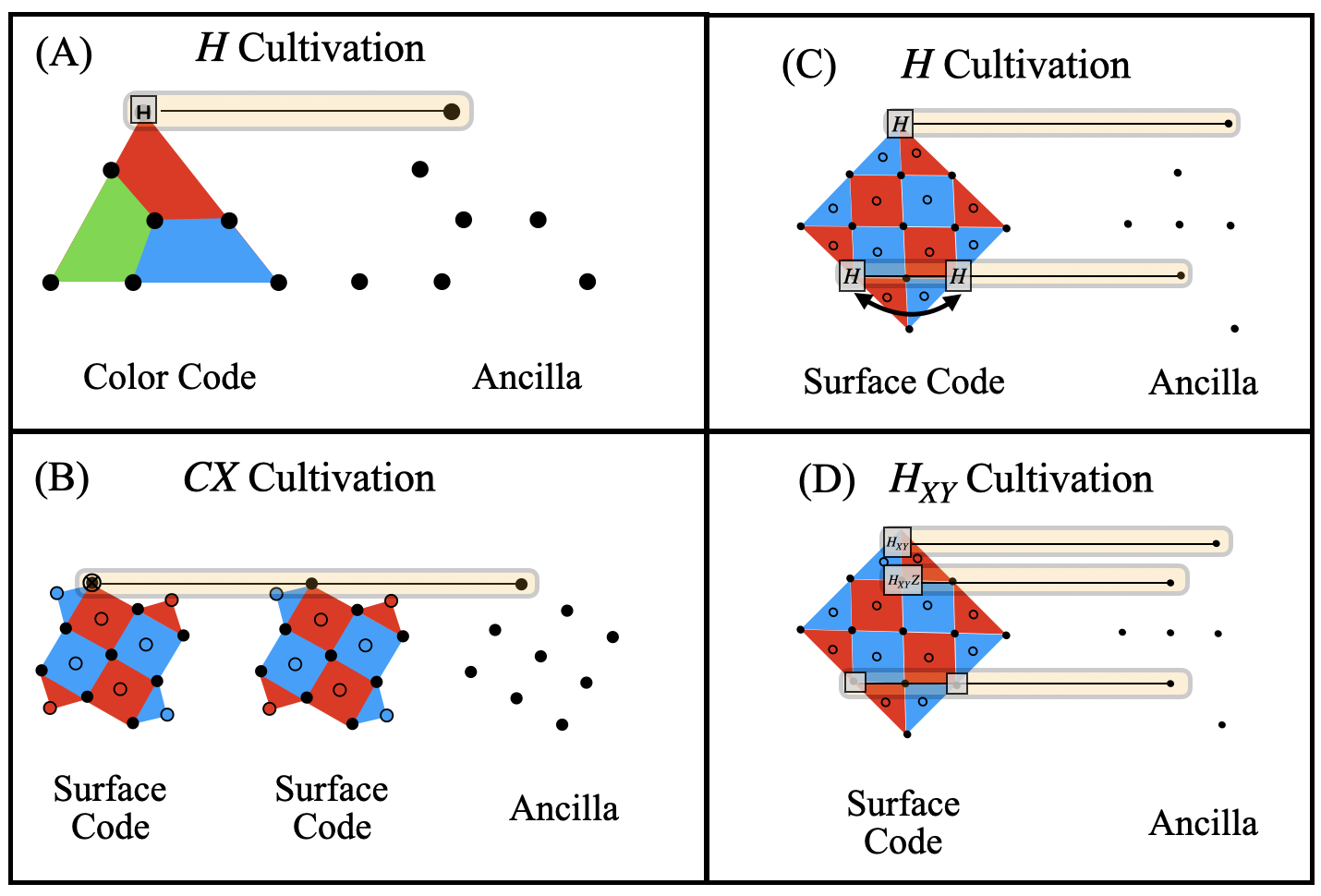}  
    
    \caption{{\textbf{$H$,$H_{XY}$ and $CX$ cultivation} -- }
    \label{fig:cultivation_digram}
    Schematic diagrams of $H$,$H_{XY}$ and $CX$ cultivation. Black dots label the data qubits, and orange rectangles label representatives of controlled transversal gates (shown in full in Fig~\ref{fig:transversal_operations}). (A) H cultivation on 2d triangular color code. A GHZ ancilla measures the H operator by transversal application of the controlled-Hadamard gate.  (B) $CX$ cultivation on a pair of $d=3$ rotated surface codes. Blue and red dots label ancillas. The $CX$ operator is measured on a pair of surface codes by transversal application of the CCX (Toffoli) gate with using an ancilla as control. (C) $H$ cultivation on the unrotated surface code, implemented using a controlled transversal $H$ on all the qubits, followed by controlled SWAP along the diagonal. (D) $H_{XY}$ cultivation on the unrotated surface code, applied using alternating controlled $H_{XY}$ and ${H}_{XY}Z$ gates along the diagonal and controlled $CZ$ gates across the diagonal.}
\end{figure}
We propose three cultivation protocols, each based on a distinct transversal operation (Fig~\ref{fig:transversal_operations}) and all relying on controlled-two-qubit gates (three-qubit gates), as shown in Fig.~\ref{fig:cultivation_digram}(B--D). Two of these protocols use transversal implementations of the Hadamard gate $H = (X+Z)/\sqrt{2}$ and the $H_{XY} = (X+Y)/\sqrt{2}$ gate to produce the resource states $\ket{T_H}$ and $\ket{T}$, respectively (defined below). $\ket{T_H}$ and $\ket{T}$ allow the realization of the non-Clifford gates $Y^{1/4}$ and $Z^{1/4}$, which can be interleaved to synthesize arbitrary single-qubit rotations. The two states differ only by a Clifford transformation, so either one is sufficient for universal quantum computation, but having protocols for both based on different physical gates offers additional flexibility in practical implementations.

The third protocol is based on the $\mathrm{CX}$ gate and prepares the resource state $\ket{CX}$ (defined below). This state is attractive because only two copies suffice to implement a Toffoli gate with probability $4/9$ via gate teleportation \cite{dennis2001toward,gupta2024encoding}, compared with seven copies of $\ket{T}$ in standard constructions. Since the Toffoli gate is often the dominant non-Clifford operation in fault-tolerant implementations of arithmetic, an efficient source of $\ket{CX}$ states can significantly reduce the overall magic-state cost of many important algorithms. In practice, we find that the $\mathrm{CX}$ cultivation protocol is not efficient enough to outperform the two constructions described above. Nevertheless, we include it here because its definition does not rely on any particular features of the surface code, and thus it may be of independent interest as a code-agnostic protocol applicable to other CSS codes that encodes a single qubits.

We will now describe the transversal implementations on the unrotated surface code, as shown in Fig~\ref{fig:transversal_operations}. The unrotated surface code has fold-duality as defined in \cite{chen2025efficient}; By swapping pairs of qubits across the code's diagonal, the resulting code is a dual of the original code, in the sense that $X$ stabilizers of the original code become $Z$ stabilizers of the dual code and vice-versa. The same goes for the $X/Z$ logical operators.

(Fold) Transversal Hadamard can be implemented by following the swap operation with $H$ gate on every data qubit, which swaps the $X$ and $Z$ operators back. An unrotated surface code has $d+2$ qubits on its diagonal and $n$ data qubits in total. We label the $d+2$ diagonal qubits as $a_i$, and the  $(n-d-2)/2$ pairs of off diagonal data qubits as $(b_i,c_i)$ and write the transversal decomposition as:
$$
A_{i}=\begin{cases}
H_{a_{i}} & i\le \left(d+2\right)\\
\text{SWAP}\left(b_{i},c_{i}\right)H_{b_{i}} H_{c_{i}} & i > \left(d+2\right)
\end{cases},
$$
where $\text{SWAP}(b_i,c_i)$ swaps between the qubits $b_i,c_i$ and $H_x$ is the $H$ operator on qubit labeled by $x$. The resulting $+1$ eigenstate of the logical $H$ is the following magic state:
$$\left|T_{H}\right>=\cos\left(\pi/8\right)\left|0\right>+\sin\left(\pi/8\right)\left|1\right>.$$
Similarly, the following transversal operator behaves as $H_{XY}$ in the logical basis of the surface code:

$$
A_{i}=\begin{cases}
\left(H_{XY}\right)_{a_{i}} & i\le\left(d+2\right)\land i\mod2=1\\
\left(H_{XY}Z\right)_{a_{i}} & i\le\left(d+2\right)\land i\mod2=0\\
CZ\left(b_{i},c_{i}\right) & i>\left(d+2\right),
\end{cases}
$$
where the subscript again labels the single qubit gate's support, and $CZ(b_i,c_i)$ is the controlled-$Z$ gate between qubits $b_i, c_i$. The details of this transformation are described in Appendix~\ref{app:transversal_HXY}. The $+1$ eigenstate of $H_{XY}$ is:
$$
\left|T\right>=\left[\left|0\right>+e^{i\pi/4}\left|1\right>\right]/\sqrt{2}
$$

\begin{figure*}[t] 
    \centering

    \includegraphics[width=0.65\textwidth,trim = {0 130 0 0}]{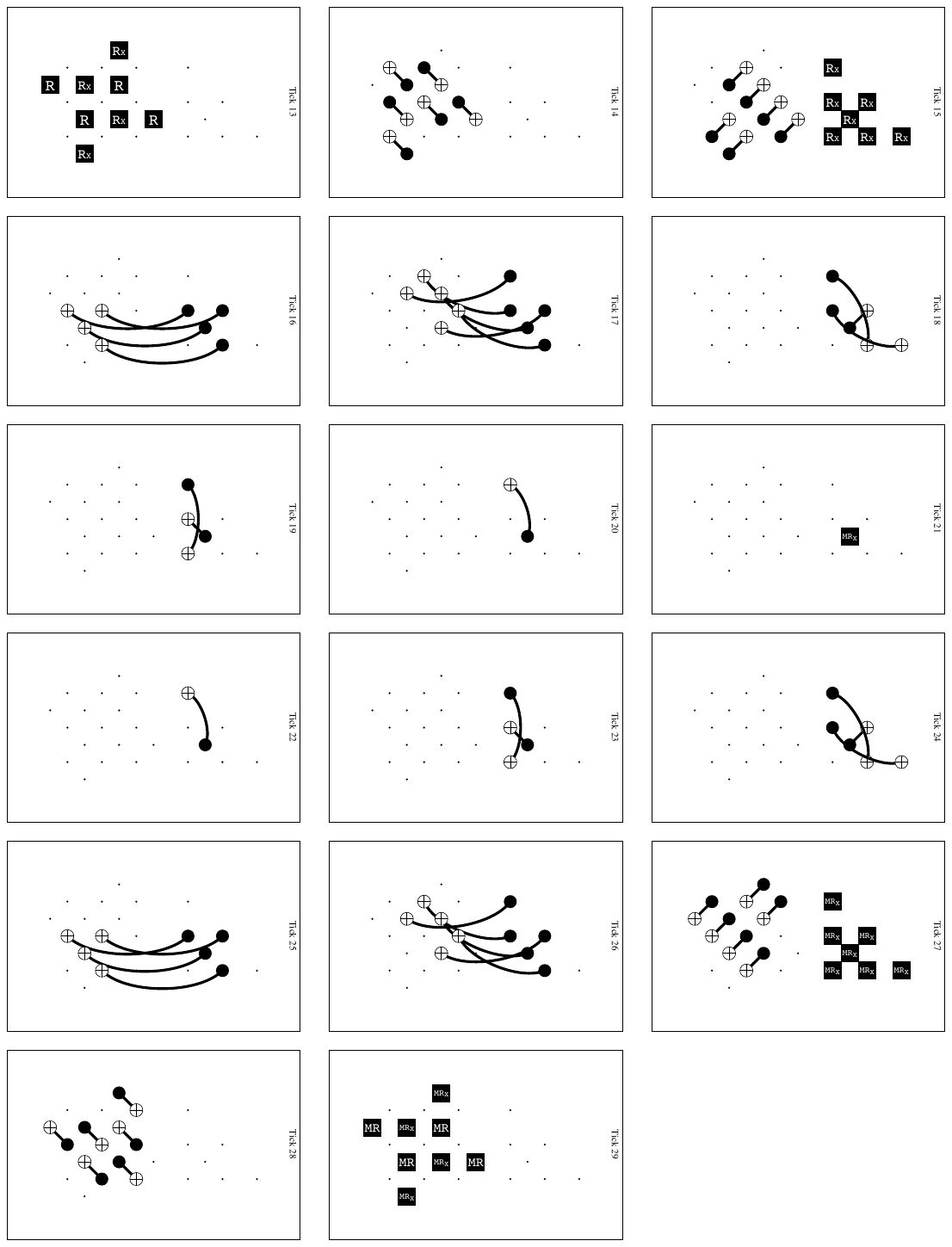} 
    \caption{\textbf{Circuit measuring the $X$ operator on a rotated surface code mid SE cycle} --
    \label{fig:transversal_x_measurement}
    A \textit{double-check} of the $X$ operator on a rotated surface code with $d_1=3$ mid SE cycle. From top to bottom, left to right, each panel represent a time slice of the circuit (labeled with tick index). For non-Clifford simulation, the $CX$'s in ticks (16,17,25,26) were replaced with $CCZ$, $CH_{\text{XY}}$ and $CH_{\text{XY}}Z$}.
    \href{https://algassert.com/crumble#circuit=Q(0,2)0;Q(1,1)1;Q(1,3)2;Q(1,5)3;Q(2,2)4;Q(2,4)5;Q(2,6)6;Q(3,1)7;Q(3,3)8;Q(3,5)9;Q(4,0)10;Q(4,2)11;Q(4,4)12;Q(5,1)13;Q(5,3)14;Q(5,5)15;Q(6,4)16;Q(8,1)17;Q(8,3)18;Q(8,5)19;Q(9,4)20;Q(10,3)21;Q(10,5)22;Q(12,5)23;R_0_5_11_16;RX_4_6_10_12;TICK;CX_2_5_7_11_14_16_4_1_6_3_12_8;TICK;CX_3_5_8_11_15_16_4_7_6_9_12_14;RX_17_18_19_20_21_22_23;TICK;CX_18_2_20_5_21_8_22_9;TICK;CX_17_4_18_7_20_11_21_12_22_14;TICK;CX_20_21_17_22_18_23;TICK;CX_20_18_17_19;TICK;CX_20_17;TICK;MRX_20;DT(9,4,0)rec[-1];TICK;CX_20_17;TICK;CX_20_18_17_19;TICK;CX_20_21_17_22_18_23;TICK;CX_18_2_20_5_21_8_22_9;TICK;CX_17_4_18_7_20_11_21_12_22_14;TICK;CX_1_0_8_5_13_11_4_2_10_7_12_9;MRX_17_18_19_20_21_22_23;DT(8,1,1)rec[-7];DT(8,3,1)rec[-6];DT(8,5,1)rec[-5];DT(9,4,1)rec[-4];DT(10,3,1)rec[-3];DT(10,5,1)rec[-2];DT(12,5,1)rec[-1];TICK;CX_2_0_9_5_14_11_4_8_10_13_12_15;TICK;MR_0_5_11_16;MRX_4_6_10_12;DT(0,2,2)rec[-8];DT(2,4,2)rec[-7];DT(4,2,2)rec[-6];DT(6,4,2)rec[-5];DT(2,2,2)rec[-4];DT(2,6,2)rec[-3];DT(4,0,2)rec[-2];DT(4,4,2)rec[-1]}{Click here to view in Crumble}
\end{figure*}

The same ideas can be used on the rotated surface code, although it is not fold-dual. The rotated surface code requires fewer physical qubits than the unrotated layout to reach similar logical error rates, so it is generally the more resource-efficient choice. As described in Ref.~\cite{chen2024transversal}, halfway through the stabilizer extraction (SE) cycle of the rotated surface code, the stabilizer group closely matches that of an unrotated surface code with the same distance. This mid-cycle equivalence allows us to implement both fold-transversal operators defined above for the \emph{unrotated} surface code directly on a \emph{rotated} surface code by applying them at this intermediate point in the stabilizer circuit, thereby combining the transversal structure of the unrotated geometry with the qubit savings of the rotated one. After the projection, the measurement cycle is completed and we measure the stabilizers and treat them as an additional ancillary checks. Fig~\ref{fig:transversal_x_measurement} shows the \textit{double-checking} circuit implemented this way on the unrotated surface code. 

Lastly, CX cultivation is based on the observation that logical $\mathrm{CX}$ is transversal for any CSS code: applying physical $\mathrm{CX}$ gates between corresponding data qubits of two code blocks implements a logical $\mathrm{CX}$ between the encoded qubits. Explicitly, if we label each pair of corresponding physical qubits from the two surface-code patches by $(a_i,b_i)$ for $1 \le i \le n$, and we can define the transversal operator $A_i = \mathrm{CX}(a_i,b_i)$. The $+1$ eigenspace of logical $\mathrm{CX}$ contains multiple logical states, but by initializing the two surface codes in the logical state $\ket{+,0}$ and projecting onto this eigenspace, the joint logical state is prepared in :
\[
    \ket{\mathrm{CX}} = \ket{0,+} + \ket{0,-} + \ket{1,+},
\]
which we use as the entangled resource state for Toffoli implementation\cite{dennis2001toward, gupta2024encoding}.

\subsection{Phase Kickback and Double Checking}
\label{sec:phase-kickback-and-double-checking}

For completeness, we will define in detail two different methods for state projection obeying Eqs~\ref{eq:logical_scaling_phase_kickback},\ref{eq:logical_scaling_double_checking}: phase kickback and \textit{double-checking}. For phase kickback we initialize a GHZ ancilla by expanding a $\left|+\right>$ state from one qubit using $CX$ gates. Then, a controlled transversal gate is applied with the GHZ state as control, as seen in Fig.~\ref{fig:phase_kickback}.

If the logical qubit is in some superposition of the two eigenstates $\left|\psi\right>=\alpha\left|A\right>+\beta\left|\bar{A}\right>$, the application of the transversal operation conditioned on the GHZ ancilla results in the following state:

{\footnotesize
\begin{align}
\left(\prod_i CA_{i}\right)
\left[\frac{\left|0\right>^{\otimes n}+\left|1\right>^{\otimes n}}{\sqrt{2}}\right]\left|\psi\right>&=\alpha\left[\left|0\right>^{\otimes n}+\left|1\right>^{\otimes n}\right]\left|A\right>\nonumber \\ 
&+\beta\left[\left|0\right>^{\otimes n}-\left|1\right>^{\otimes n}\right]\left|\bar{A}\right>,\nonumber
\end{align} }
Where $CA_i$ is a transversal $A_i$ gate conditioned on the $i$'th ancilla in the GHZ state. Inverting the GHZ generation and then measuring its qubits effectively measures the GHZ state and can identify a single bit flip error \cite{chamberland2020very}. 

Note that if the transversal operation did not act trivially on the stabilizers of the code, it would generate additional entanglement between the code state and the ancilla. In such a case, the superposition of the ancilla would be destroyed, leaving it in a mixed state over $\ket{0}^{\otimes n}$ and $\ket{1}^{\otimes n}$. A subsequent measurement would then fail to reveal any information about the $\ket{A}$ state. To avoid this, our initialization sets all stabilizers to the $+1$ eigenspace, making $A$'s automorphism of the stabilizer group trivial. 

In the \textit{double-checking} protocol, first introduced in \cite{gidney2024magic}, all the ancilla qubits are initialized in the $\left|+\right>$ state. The controlled transversal operation is applied, which puts the ancilla in the following superposition:
{
\begin{align}
\left(\prod CA_{i}\right)\left|+\right>^{\otimes n}\otimes\left|\psi\right>&=\alpha\prod\left(\left|0\right>+\left|1\right>A_{i}\right)\left|A\right>\nonumber\\
&+\beta\prod\left(\left|0\right>+\left|1\right>A_{i}\right)\left|\bar{A}\right>
\end{align}
}
Notice that $\prod\left(\left|0\right>+\left|1\right>A_{i}\right)\left|A\right>$ is a $+1$ eigenstate of the $X^{\otimes n}$ operator on the ancilla, therefore  eigenvalue of $A$ is encoded in the parity of $X^{\otimes n}$. We measure this parity using a series of $CX$ gates that map by conjugation $X^{\otimes n}$ to $X$ operator of one of the ancillas. Measuring this ancilla in the $X$ basis measures the value of $A$. The number of checks is doubled by running the circuit in reverse and measuring all of the ancilla qubits in the $X$ basis, see Fig \ref{fig:phase_kickback_double_checking}.

\begin{figure}
\[
\Qcircuit @C=0.25em @R=.65em {
&\lstick{\ket{+}} & \ctrl{3} &\qw&\qw   &\ctrl{2} & \ctrl{1}& \measureD{X} & \ctrl{1}& \ctrl{2}& \ctrl{3} &\qw&\qw   & \measureD{X}  \\
&\lstick{\ket{+}} & \qw& \ctrl{3}& \qw  &\qw      & \targ   & \qw           & \targ   & \qw     & \qw& \ctrl{3}& \qw  & \measureD{X}  \\
&\lstick{\ket{+}} & \qw&\qw&\ctrl{3}    &\targ    & \qw     & \qw           & \qw     & \targ   & \qw&\qw&\ctrl{3}    & \measureD{X}  \\
&\qw              &  \gate{A_1} &\qw&\qw  &\qw      & \qw     & \qw           & \qw     & \qw     &  \gate{A_1} &\qw&\qw  &\qw \\
&\qw              & \qw& \gate{A_2}&\qw   &\qw      & \qw     & \qw           & \qw     & \qw     & \qw& \gate{A_2}&\qw   &\qw \\
&\qw              &  \qw&\qw& \gate{A_3}  &\qw      & \qw     & \qw           & \qw     & \qw     &  \qw&\qw& \gate{A_3}  &\qw  
\inputgroupv{4}{6}{.8em}{1.9em}{\ket{\psi}}\\
}
\]
    \caption{{\textbf{Schematic description of double-checking} -- }
    \label{fig:phase_kickback_double_checking}
    The top 3 Qubits represent the ancilla qubits, while the bottom 3 Qubits represent the data qubits of the surface code (or codes). The circuit projects $\ket{\psi}$  to the $+1$ eigenvector of $A$ when all measurement outcomes are $+1$. The state is discarded if any measurement outcome is $-1$ using \textit{double-checking}. } 
\end{figure}

\subsection{Code Expansion}

An important optimization of the cultivation protocol is to gradually increase the code size \cite{goto2016minimizing, gidney2024magic}. The protocol described so far aborts following most physical errors, which makes the acceptance rate highly sensitive to the number of active qubits. By activating some of the qubits only in the later stages of the protocol, we can substantially increase the acceptance rate.

Concretely, we first run cultivation on a small $d=3$ code, and then use a short circuit to transform the code into a larger $d=5$ surface code. As long as the expansion circuit, together with the final projection step, continues to obey Eqs.~\eqref{eq:logical_scaling_phase_kickback} and~\eqref{eq:logical_scaling_double_checking}, the output fidelity retains the same asymptotic scaling. We note that this argument does not work for CX cultivation, as explained in Appendix~\ref{SI:cx_cultivation}. 

We will follow the convention of Refs.~\cite{gidney2024magic, chen2025efficient}, and label each protocol by $d_1$, the code distance used at the final projection step and therefore the distance that determines the asymptotic logical scaling. 

Because cultivation relies on post-selection, it can be implemented at relatively small code distances, whereas the subsequent fault-tolerant computation must run without post-selection and therefore requires a higher distance. Following Ref.~\cite{gidney2024magic}, we bridge these two regimes by adding a final code-expansion stage that moves the cultivated state onto a larger code suitable for standard error correction, with only limited additional post-selection. We label that final distance $d_2$. After the last cultivation measurement, each surface-code patch is expanded \cite{goto2014step,gidney2023cleaner,gidney2024magic} to a distance-$d_2 = 11$ code and subjected to 10 further rounds of stabilizer measurements\footnote{In  Appendix \ref{SI:post_expansion_rounds} we show that our protocol is insensitive to these exact numbers, which should only be understood as a convention}.The expansion is performed by initializing each newly added qubit in the eigenbasis of any weight-2 stabilizer or logical operator it extends, so that all stabilizer and logical eigenvalues remain deterministic during the growth process \cite{gidney2023cleaner}. In our simulations we expanded (un)rotated surface codes to larger (un)rotated surface codes. however, since rotated and unrotated surface codes can both be embedded into sufficiently large codes of either type, the same procedure can in principle be used to expand an unrotated code into a rotated one, which would be more qubit efficient. 

At this stage we no longer post-select on every detected syndrome; instead, we follow the soft-decoding strategy of Refs.~\cite{bombin2024fault, gidney2023yoked, gidney2024magic}. The decoder finds the most probable error resulting in either a logical phase error or no error, and their respective probabilities $p_Z, p_I$.  These probabilities define the complementary gap $\mathsf{gap}=\log{(p_Z/p_I)}$. By imposing a threshold on this gap, we obtain a family of protocols that trade acceptance rate against output fidelity, and in our results we sweep this threshold to map out the full acceptance–fidelity frontier.

\subsection{Protocol}

\begin{figure}
    \centering
    \includegraphics[width=0.42\textwidth,
    trim={25mm 25mm 25mm 25mm}]{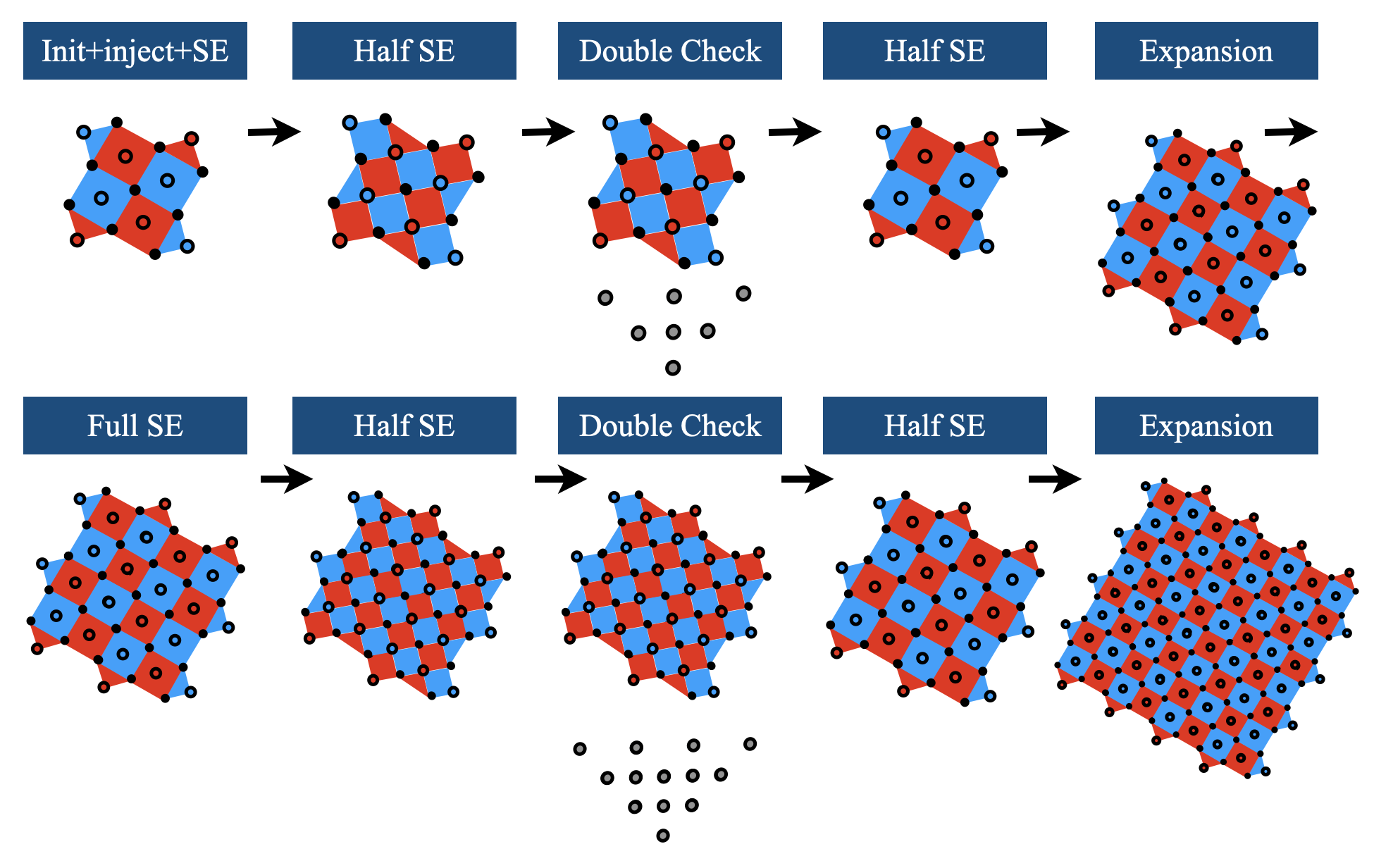}  
        \caption{{\textbf{Full protocol for $d_1=5$ rotated surface code} -- }
    \label{fig:full_protocol}
Step-by-step illustration of the cultivation protocol. Filled black dots denote data qubits. Hollow red (blue) dots denote ancilla qubits used for syndrome extraction of $X$ ($Z$) stabilizers. Hollow gray dots denote ancilla qubits used for double-checking. \textit{SE} refers to the syndrome-extraction circuit, and \textit{half SE} refers to two of the four CX gates required for a surface-code SE round.The red and blue polygons indicate the state of the stabilizer group after each labeled step \cite{chen2024transversal}.
    }
\end{figure}

We now describe the step-by-step structure of our cultivation protocol. The $H$ and $H_{XY}$ schemes follow the same sequence of operations, differing only in the specific transversal operator $A$ and in the details of the injection step. The protocol proceeds as follows:

\begin{enumerate}
    \item \textbf{Initialization.} \\
    A distance-$3$ surface code is initialized in the $+1$ eigenstate of the logical $Z$ operator and in the $+1$ eigenspace of all stabilizers. This can be achieved by measuring all stabilizers once and applying physical corrections conditioned on the outcomes.

    \item \textbf{State injection with syndrome extraction.} \\
    A non-fault-tolerant unitary injection circuit is applied to map the encoded state to the target magic state $\ket{A}$ (see Appendix~\ref{SI:injection} for details). We then perform a single round of the standard syndrome-extraction circuit. The run is discarded if a syndrome is observed.

\item \textbf{Double-checking} \\
The logical operator $A$ is measured using the double-checking circuit described in Sec.~\ref{sec:phase-kickback-and-double-checking}. The run is aborted if any measurement yields a nontrivial outcome.

\item \textbf{Second double-checking with code expansion.} \\
Only for protocols we label as $d_1=5$, we expand the code before performing an additional projection round. We initialize the additional qubits required for a $d=5$ surface code, half in the $X$ basis and half in the $Z$ basis, and apply a short expansion circuit that maps the $d=3$ code into a $d=5$ code while preserving the $+1$ eigenvalues of the stabilizers. We run a single syndrome extraction round, and then apply a second \textit{double-checking} circuit. 

    \item \textbf{Code expansion and soft-decoded error correction.} \\
    The accepted state is expanded to a larger surface code. In our simulations we expand to distance $d_2=11$ and perform 10 rounds of syndrome extraction. At this stage we do not post-select on every detected syndrome. Instead, we apply soft decoding and accept the run based on a threshold on the decoder confidence, quantified by the complementary gap. 
\end{enumerate}
We present this in full as psudo-code in Appendix \ref{SI:psudocode}. For CX cultivation, we use a slightly different protocol, see Appendix~\ref{SI:cx_cultivation}.

\section{Simulation}

We next turn to numerical simulations. For small code distances, we verified the protocol’s correctness using full state-vector simulations. Specifically, we simulated both $H$ and $H_{XY}$ with rotated and unrotated codes with $d_1=3$, and for $CX$ cultivation we simulated a representative example with $d_1=2$. 

To estimate the logical performance of our protocols, we introduce noise into the simulations controlled by a single parameter $p$. After every single-qubit (two-qubit) gate, we apply a single-qubit (two-qubit) depolarizing channel with probability $p$. After each three-qubit operation, we apply a three-qubit depolarizing channel with probability $3p$, together with additional two-qubit depolarizing channels between the ancilla control and each data-qubit target, each occurring with probability $p$. The full noise model is specified explicitly in Appendix~\ref{SI:Noise Model}. 

\begin{figure} 
    \centering
    \includegraphics[width=0.50\textwidth,
    trim={20mm 20mm 20mm 20mm}]{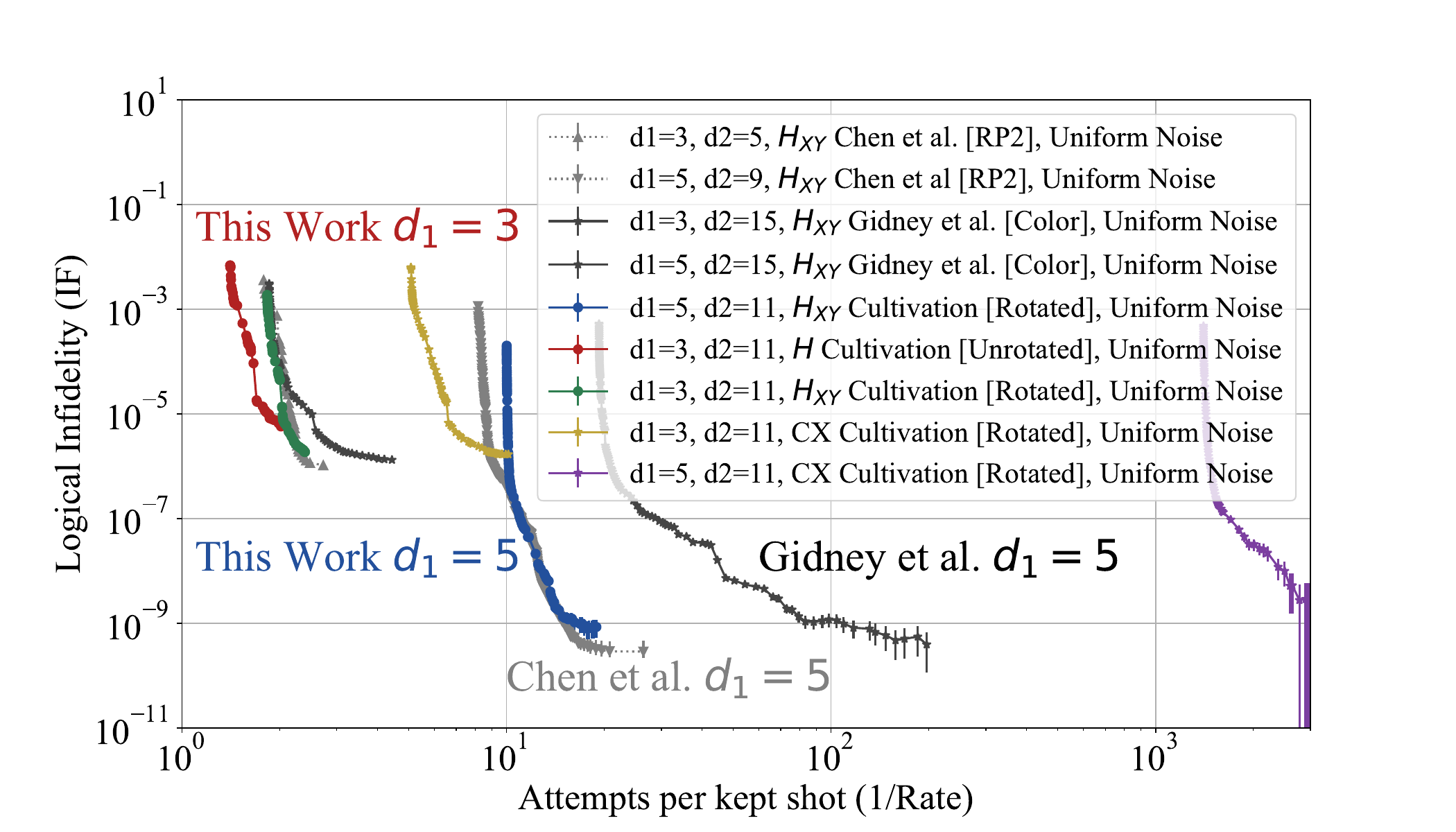}  
    
    \caption{{\textbf{Logical error rate of cultivation with $d_2$ expansion - Uniform noise model} -- }
    \label{fig:cultivation_and_expansion}
    Logical infidelity, and acceptance rate of our protocol compared with previous protocols \cite{gidney2024magic, chen2025efficient}. The physical noise parameter is $p=10^{-3}$ using uniform noise model. The color code from \cite{gidney2024magic} is expanded to a variant of the surface code, while the rest are expanded to a standard surface codes, each with their labeled distance $d_2$. The different points label various thresholds for the complementary gap post-selection. Error bars represent a single standard deviation.   We included the $d_1=5$ $CX$ case although it has large error bars, giving  only an approximate bound on its error rate. Note that Ref.~\cite{chen2025efficient} uses different values of $d_2$, which we argue in Appendix~\ref{SI:post_expansion_rounds} has a negligible effect on this figure.}
\end{figure}

While the computational cost of full state-vector simulation allowed us to simulate the protocol up to distance $d_1 = 3$ without expansion, it became prohibitive to directly estimate the logical error rate after the final expansion to distance $d_2$, or even after the shorter expansion to $d_1 = 5$. To access this larger-scale regime, we therefore follow Ref.~\cite{gidney2024magic} and employ a closely related circuit that replaces the non-Clifford gates with Clifford gates and projects a different logical operator. This enables efficient sampling via Clifford simulation \cite{gottesman1998heisenberg, aaronson2004improved,gidney2021stim, gidney2024magic}. Concretely, in the Clifford simulation \cite{aaronson2004improved,gidney2021stim} we replace the non-Clifford $CA_i$ gates with combinations of $CX$ designed to measure the logical $X$ operators. We introduce noise as depolarizing channels as defined in Appendix~\ref{SI:Noise Model}. To ensure that our simulation approximates the noise in the system, we apply a uniform depolarizing noise on the support of the \textbf{non}-Clifford gate, which sometimes includes 3-qubit depolarizing errors, see Appendix~\ref{3qubit_gate_compilation}. The results of these simulations are shown in Fig.~\ref{fig:cultivation_and_expansion}.

Similarly to Ref.~\cite{gidney2024magic}, we observe good agreement in both absolute fidelity and scaling between the full-vector simulations and the simplified Clifford simulations, with only a small factor difference before the expansion step. See Appendix~\ref{SI:full_vector_simulation} for details. Owing to its smaller qubit footprint, the rotated surface code provides the best overall fidelity, even though its circuit is longer. When idling errors are absent, the shorter circuit of the unrotated surface code performs better, as demonstrated in the next section.

Finally, we compare our protocol with color-code-based magic state distillation \cite{chamberland2020very, gidney2024magic}. Although color-code-based protocols have an initial advantage due to a smaller qubit footprint, our protocol is ultimately superior. The expansion step in our protocol is substantially more efficient than the costly grafting procedure required for color codes. Grafting requires a long idling step with high rates of post selection, necessary to convert from color code to surface code. Our protocol skips this idling step entirely, since we simply expand to a larger surface code. This results in an order-of-magnitude higher distillation rate for representative parameters of $d_1=5, d_2=11$, which are easily comparable with previous work \cite{gidney2024magic}. Our protocol is able to reach $10^{-9}$ infidelity with the same physical error rate $p=10^{-3}$ and $1/5$ the number of shots.

\subsection{Efficient Cultivation with long idling time}

To enable direct comparison with previous schemes, we adopt the uniform-noise model described in Ref.~\cite{gidney2024magic}. 
This model treats idling as equally harmful as a two-qubit gate, which can be a reasonable approximation for certain superconducting transmon devices~\cite{hashim2020randomized, foxen2020demonstrating}.
However, some quantum devices offer significantly better idling fidelities compared to their 1Q and 2Q gates. For example, Cold atom devices have exceptionally long $T_1$ times while not populating the Rydberg state, and their dephasing time $T_\varphi$ can be prolonged with dynamical decoupling, or by shelving in clock states~\cite{manetsch2025tweezerarray6100highly, bluvstein2025architectural}. This bias between gates and idling can also be achieved in ions and superconducting cavities~\cite{ryananderson2022ions,Ryan-Anderson2021ionsNoise2,leblond2025logicalerrorratessurface} and possible with spins in solids, like \cite{steinacker2024300}.

\begin{figure} 
    \centering
    \includegraphics[width=0.50\textwidth,
    trim={20mm 20mm 20mm 20mm}]{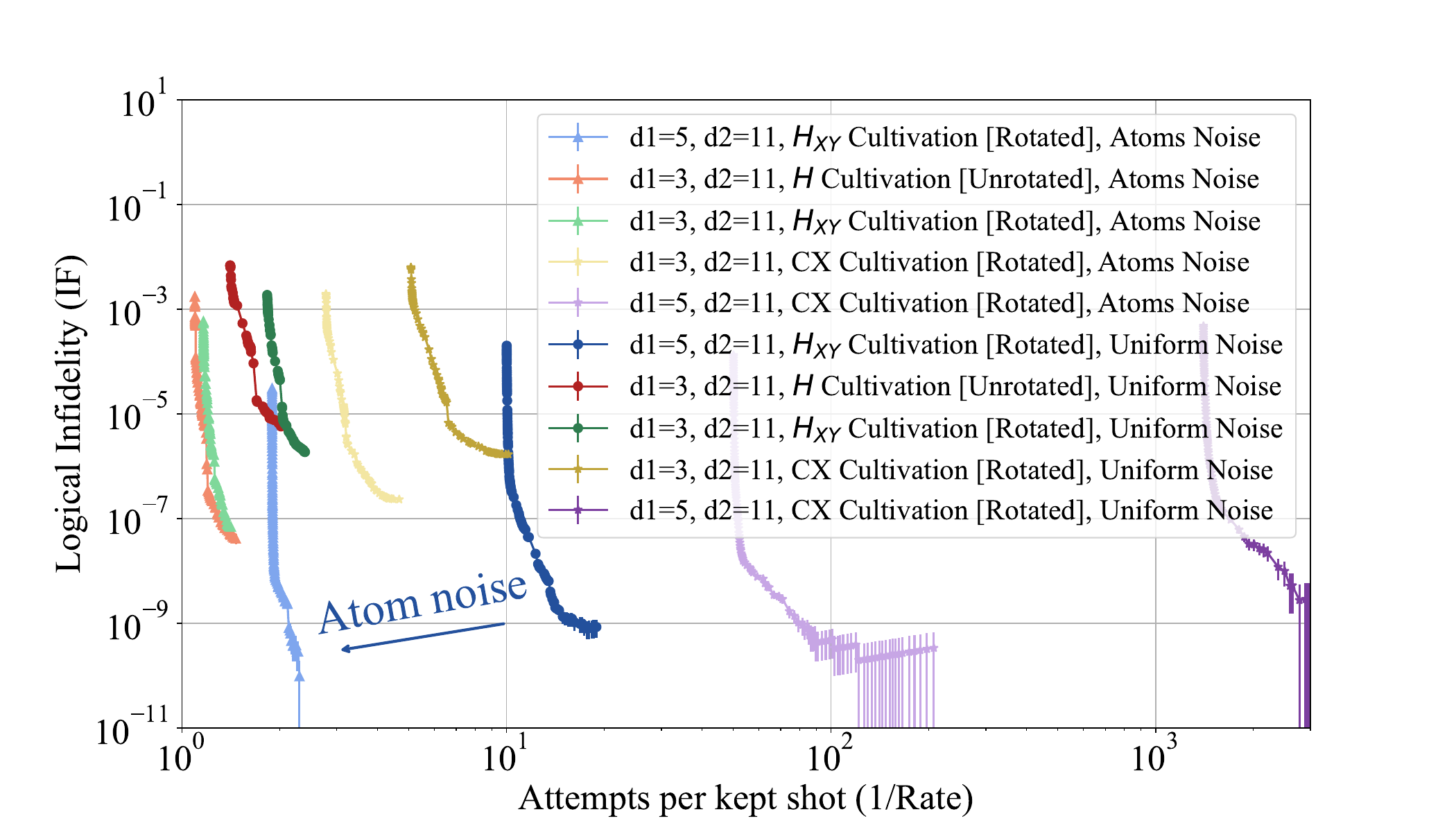}  
    
    \caption{{\textbf{Logical error rate of cultivation with $d_2$ expansion - Atom noise model} -- }
    \label{fig:neutral_atoms_noise_Model}     Logical infidelity and acceptance rate of surface code $CX, H_{XY}$ and $H$ cultivation with Uniform and Atom noise model. In the Atom noise model, idling doesn't introduce any additional error, which approximate the effect of dynamical decoupling or shelving in clock-states. The physical error rate is $p=10^{-3}$, and the full noise model is defined in S.I. The different points label various thresholds for the complementary gap post-selection. For practical application that utilize synthillation, achieving $10^{-7}$ infidelity is possible with $d_1=3$ and approximately $75\%$ rate. }
\end{figure}

We modeled this effect by defining a different noise model, which we call atom-noise model. In this model, we reduce the 1-Qubit gates error probability to $p/10$, and remove any idling noise (Appendix~\ref{SI:rydbergatoms_noise}). The results of simulations with atom-noise model are shown in Fig~\ref{fig:neutral_atoms_noise_Model}.

This offers a significant advantage in practice. In a recent estimate for the resources required to break RSA 2048 using superconducting qubits, roughly $20\%$ of the qubits were designated for magic state generation \cite{gidney2025factor}. Using color code cultivation and targeting $10^{-7}$ infidelity, the acceptance rate is approximately $5\%$. With $d_1=3$ $H_{\text{XY}}$ cultivation we have roughly $75\%$ acceptance rate with fewer qubits and roughly half the number of rounds. In total, this results in over an order of magnitude reduction in qubit-rounds cost.

 Still, in a resource estimate designed specifically for cold atoms \cite{zhou2025resource}, magic-state cultivation occupies half the atoms during addition (where most of the Toffoli gates are consumed). Our short circuit and $75\%$ acceptance rate together would reduce the qubit-rounds assumed in \cite{zhou2025resource} by over a factor of $\times 20$. This would directly reduce the qubit count during addition by over $6$ million qubit (or roughly $40\%$). 

\section{Erasure Qubits and Loss}

Erasure qubits \cite{bartolucci2023fusion,kubica2023erasure,EruasePatent,levine2024demonstrating,wu2022erasure,gu2024optimizing,gu2024optimizing} are a type of qubits with a specific noise mechanism that can be directly identified. In superconducting qubits, energy relaxation events can be identified by encoding a single qubit in the single excitation subspace of two transmons \cite{kubica2023erasure, levine2024demonstrating} or cavities \cite{chou2024superconducting,koottandavida2024erasure}, and measuring the zero excitation state. In Rydberg atoms \cite{ma2023high,scholl2023erasure,wu2022erasure}, during a two qubit gate the population in the Rydberg state decays to many states, most of them outside of the computational subspace. Observing the population in these states identifies this decay. Similarly, loss in Rydberg atoms can be detected through destructive measurements \cite{baranes2025leveraging,bluvstein2025architectural}. In the context of cultivation, any technique that identifies loss by the end of the protocol effectively functions as erasure detection \cite{baranes2025leveraging}.

The erasure rate doesn't effect the logical fidelity \cite{jacoby2025magic} as long as it can be detected. In this limit, only the acceptance rate is affected by erasure.  Since most Pauli errors are identifiable to begin with \cite{jacoby2025magic}, the effect of erasure $e$ is approximately to reduce the acceptance rate to the rate of the same circuit with only Pauli noise with rate $p=e$. See Appendix~\ref{SI:erasure} for our analysis and \cite{jacoby2025magic}, for a detailed investigation of this trade off.

\begin{figure}
    \centering

        \includegraphics[width=0.50\textwidth,
    trim={20mm 20mm 20mm 20mm}]{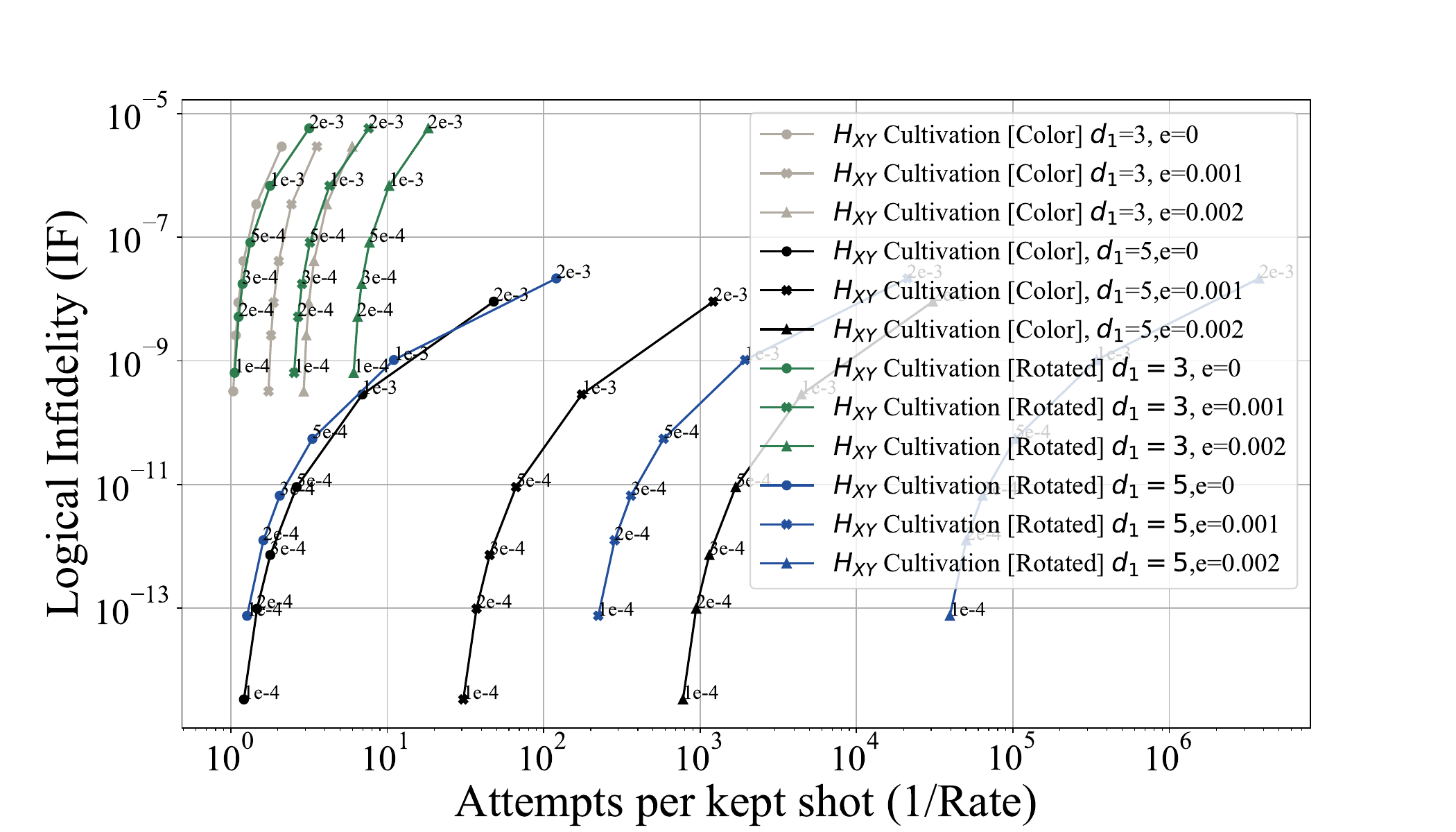}
    \caption{\textbf{Logical error rate and Acceptance rate of erasure Qubits} --
    \label{fig:erasure}
Logical error rate and acceptance rate of erasure qubits used for the initialization step of cultivation. The points are labeled by the residual Pauli error rate $p$. As mentioned in the main text, the logical error rate does not depend on the erasure rate but only on the residual Pauli error rate. The logical infidelity is estimated by full search on the error graph up to weight 5, and matches sampling for values $p$'s where sampling is possible. Importantly, the color-code expansion step is considerably more expensive than for the surface code, but isn't shown in this plot. Still, from this graph we can extract the tradeoff when converting between different erasure and Pauli rates. For example, if $p=10^{-3}$ can be converted to $e=2 \times 10^{-3}$ and $p=10^{-4}$, it would reduce the acceptance rate of $d_1=3$ cultivation by roughly a factor of $5$, but improve the fidelity by over $10^3$. 
}
    
\end{figure}
We estimated the effect of erasure on cultivation by evaluating its effect on acceptance rate without the final $d_2$ expansion step in Fig. \ref{fig:erasure}. In appendix~\ref{SI:erasure}, we provide an explanation of the estimation process for assessing the impact of erasure on complementary gap decoding. We find that the overhead of adding  $e=2\times 10^{-3}$ erasure probability per operation for $d_1=3$  surface code cultivation roughly reduces the acceptance rate by a factor of $5$. If in this regime one can achieve a meaningful bias, resulting in $p=10^{-4}$ , it can reach infidelity of $10^{-9}$ with the same rate as $d_1=5, p=10^{-3}$ cultivation but with significantly fewer qubits and a shorter protocol. Such a level of bias is realistic for certain trapped-atom species \cite{wu2022erasure} and for superconducting qubits \cite{levine2024demonstrating}, although the latter would require some degree of long-range connectivity.

If the bias is small, which is the case for some trapped atom species \cite{bluvstein2025architectural} where loss represent about $50\%$ of gate fidelity, cultivation is still improved by loss detection. The rate would be the same as we described before, but the final fidelity would improve by about $2^{d_1}$, which as seen in Fig~\ref{fig:erasure} is between $8-50$ depending on the specific protocol.

If the bias can reach $e=10^{-3}$ and $p=10^{-4}$ with only local connectivity, such as for dual-rail superconducting qubits, it would be possible to achieve $p_L\approx 10^{-15}$ once every 200 attempts with $d_1=5$  color code cultivation of \cite{gidney2024magic}, if we include post selection on the complementary gap. At this scale, the qubit-cycles count of cultivation is $1.5 \times 10^5$ \footnote{This is the qubit-cycles count of H cultivation with $p_0=10^{-3}$ as reported in \cite{gidney2024magic}.  In the erasure case, we assume the erasure overhead of Pauli and erasure are the same, and the additional $p=10^{-4}$ Pauli errors have minimal additional overhead \cite{jacoby2025magic}. Since gap-estimation of erasure scales better than Pauli noise, this result should be understood as the bounds on qubit-cycles count.}, which is about $20$ times less than the $\left(\text{15-to-1}\right)_{13,5,5}^{6}\times\left(\text{15-to-1}\right)_{29,11,13}^{6}$ scheme of \cite{litinski2019magic} using non-erasure qubits.

Erasure offers an additional interesting opportunity for cultivation, which is to reach algorithmic interesting infidelities with only a $d_1=2$ unrotated surface code and a single GHZ projection. This protocol requires only 9 physical qubits. With this setup and a residual Pauli error rate of $p=10^{-4}$, cultivation can achieve a state infidelity of $10^{-6}$, which can then be combined with synthillation~\cite{Jones_2013} to reach algorithmically relevant fidelity. For more information, see Appendix~\ref{SI:small_erasure}

\section{Discussion}

In this work, we demonstrated how cultivation can be implemented directly on the surface code using non-local connection. Although the initialization step of cultivation on the surface code has a slightly lower rate compared with color-code cultivation, this effect is compensated by the simple and high rate expansion step of the surface code to higher distances.  This is clearly superior in architectures with all-to-all connectivity such as cold atoms and trapped ions. Even with architectures exhibiting limited long-range connectivity (e.g., superconducting qubits \cite{he2025extractors}), compiling non-local interactions will be quite efficient, given their small number. 

The acceptance rate could be further improved by providing the decoder with additional information, as in Ref.~\cite{chen2025efficient}. In our analysis we adopt the decoder-latency assumption of Ref.~\cite{gidney2024magic}: we simulate 10 extra rounds of syndrome extraction after cultivation before declaring the state ready. For superconducting qubits this corresponds to roughly a $10\,\mu\mathrm{s}$ decoding delay. If the decoder responds faster, fewer waiting rounds are needed, consistent with Ref.~\cite{chen2025efficient}. This 10-round convention significantly affects the qubits$\times$rounds metric. In architectures with longer measurement times, decoder latency can be effectively hidden, so the same waiting rounds would not appear as an added cost in qubits$\times$rounds. Although this still slows wall-clock cultivation, it can make cultivation appear cheaper under qubits$\times$rounds. We emphasize that, provided $d_2 \ge 2d_1+1$, both the choice of $d_2$ and the number of post-expansion rounds have negligible impact on acceptance rate and output fidelity, but a large impact on qubits$\times$rounds (see Appendix~\ref{SI:post_expansion_rounds}). A more direct comparison would incorporate transport and two-qubit gate times, and would choose $d_2$ based on how the cultivated state is consumed in the target circuit. These effects are not well captured by qubits$\times$rounds, so we do not attempt that comparison here and leave it for future work.

Many quantum computing architectures have high idling fidelity. As we have shown, this gives a great benefit for magic state cultivation, both in fidelity and rate. For example, running 2048 bit RSA only requires roughly $10^{-7}$ $T$ gate infidelity \cite{gidney2025factor,zhou2025resource}, which can be achieved with $d_1=3$ cultivation on cold atoms with very minimal loss detection. Since prior work assumed cultivation was successful with such small probability, it either required idling until the magic states are ready \cite{gidney2025factor}, or significantly increasing the number of qubits \cite{zhou2025resource}. Our high rates suggest magic states can be generated highly efficiently, which would have very large effect on previous resource estimates. 

\section{Acknowledgments}
We thank Craig Gidney for suggesting the mid-cycle measurement of the $H_{XY}$ operator, and for making Stim, Sinter and Crumble \cite{gidney2021stim} open source, all of which were instrumental for this project.

The authors were made aware of two related publications posted on arXiv concurrently with this work: one from Puri’s group, which investigates fold transversal cultivation on the surface code \cite{sahay2025fold}, and another from J. Claes, which demonstrates surface code cultivation using two-qubit gates \cite{claes2025cultivating}.

We thank the staff from across the AWS Center for
Quantum Computing that enabled this project. We also
thank Simone Severini, James Hamilton, Nafea Bshara, and Peter DeSantis at AWS, for their involvement and support of the research
activities at the AWS Center for Quantum Computing.

\section{Circuits and Data}

The circuits, decoders and data are available in \cite{vaknin2025github}.
\bibliography{bib}

\clearpage

\appendix
\section*{Appendix}





\section{Transversal \texorpdfstring{$H_{XY}$}{HXY} and $H$}
\label{app:transversal_HXY}

In this appendix we explain in more detail why the transversal operator $H_{XY}$ and $H$, as defined in the main text, acts as the logical operators on the unrotated surface code and induces an automorphism of its stabilizer group. Since the stabilizer group of the rotated surface code at the middle of the syndrome extraction circuit \cite{chen2024transversal} is the same as the unrotated surface code, the excat same argument applies in the rotated case.

\begin{figure*}
    \centering
    \includegraphics[width=0.8\textwidth]{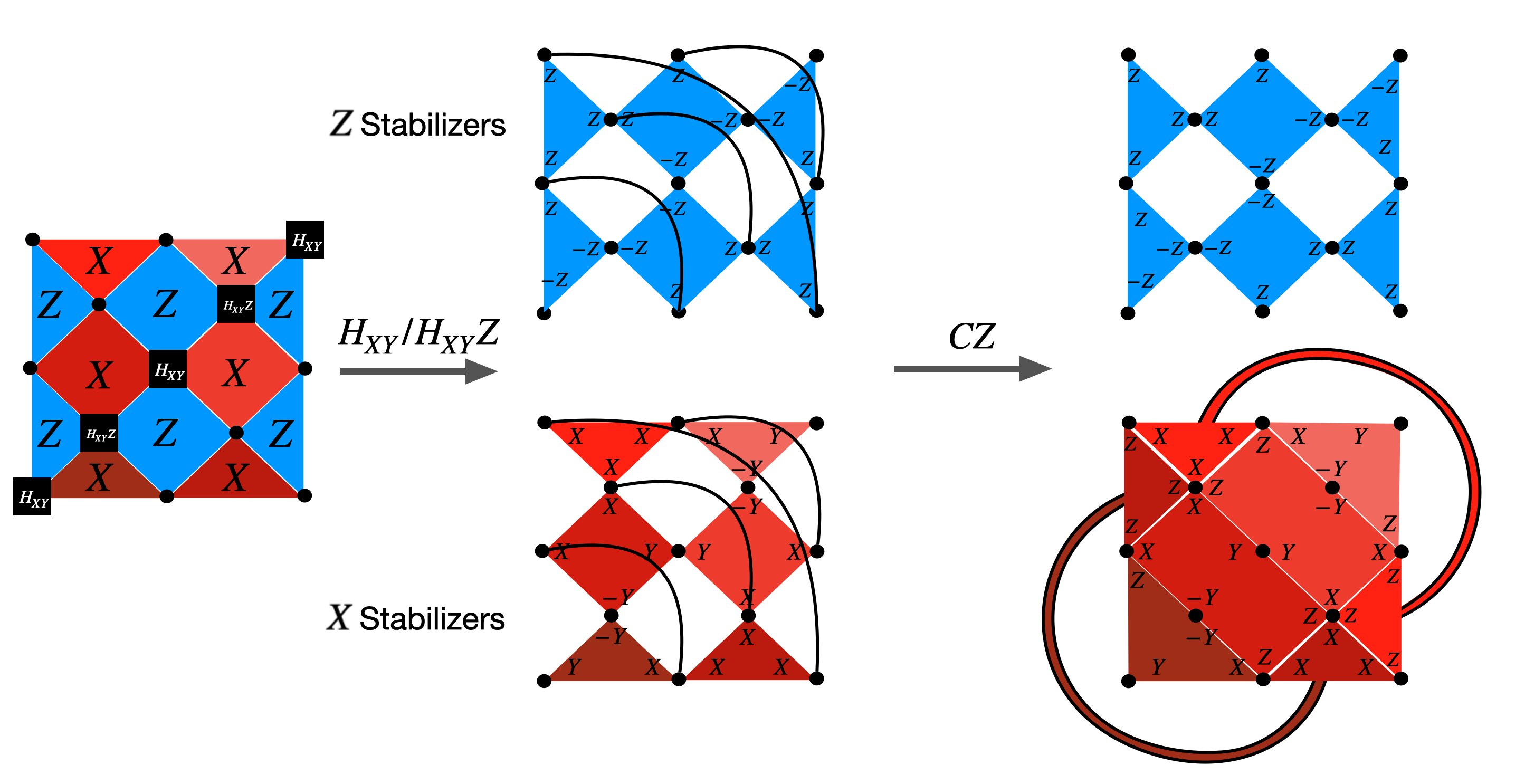}
    \caption{\textbf{Action of transversal $H_{XY}$ on the surface-code stabilizers.}
    Action of the transversal $H_{XY}$. Each panel shows the state of the stabilizer group before the operation depicted in it, and the last panel shows the final form of the stabilizers. }
    \label{fig:HXY_stab_group}
\end{figure*}

Recall that, under conjugation by $H_{XY}$ on a single qubit,  
\[
X\mapsto Y,\quad  
 Y \mapsto X,\quad  
Z\mapsto -Z,
\]
and that the two-qubit gate $CZ$ acts by conjugation as
\[
CZ\,(X\otimes I)\,CZ^\dagger = X\otimes Z,\quad
CZ\,(I\otimes X)\,CZ^\dagger = Z\otimes X,
\]
while commuting with all single-qubit $Z$ operators. We visualize these transformations in Fig~\ref{fig:HXY_stab_group}, which is the simplest way to understand the following argument.

First consider the action on the $Z$-type stabilizers. Along the diagonal, we apply alternating single-qubit gates $H_{XY}$ and $H_{XY}Z$. Each diagonal $Z$ operator therefore acquires a factor of $-1$. Since every $Z$-type stabilizer has even support on the diagonal, the overall phase from the diagonal cancels and each $Z$ stabilizer is mapped back to itself. The off-diagonal $CZ$ gates commute with all $Z$ operators, so they also preserve every $Z$-type stabilizer. In combination, the full transversal operator leaves the $Z$-stabilizer group invariant.

Next we use the fold duality of the unrotated surface code. For each $X$-type stabilizer $S_X$ there is a corresponding $Z$-type stabilizer $S_Z$ obtained by reflecting across the diagonal. Conjugation by the transversal $H_{XY}$ maps each $X$-type generator to the product $S_X S_Z$, as illustrated in Fig.~\ref{fig:HXY_stab_group}. Since the $Z$-stabilizer group is preserved, this mapping takes the $X$-stabilizers to elements of the stabilizer group and thus defines an automorphism of the full stabilizer group.

Finally, we consider the logical operators. The logical $Z_L$ string intersects the diagonal in a single qubit, so it picks up a minus sign from the action of $H_{XY}$ on that qubit, and is otherwise unchanged. Hence $Z_L \mapsto -Z_L$. The logical $X_L$ string crosses $Z_L$ at a single qubit, and on that qubit $X$ is mapped to $Y$ by $H_{XY}$. Combined with mapping $X$ terms to an $X\otimes Z$ product with their fold-dual qubit, this implies that the logical operator transforms as $X_L \mapsto i X_L Z_L = Y_L$, as required from a logical $H_{XY}$.

\begin{figure*}
    \centering
    \includegraphics[width=0.8\textwidth]{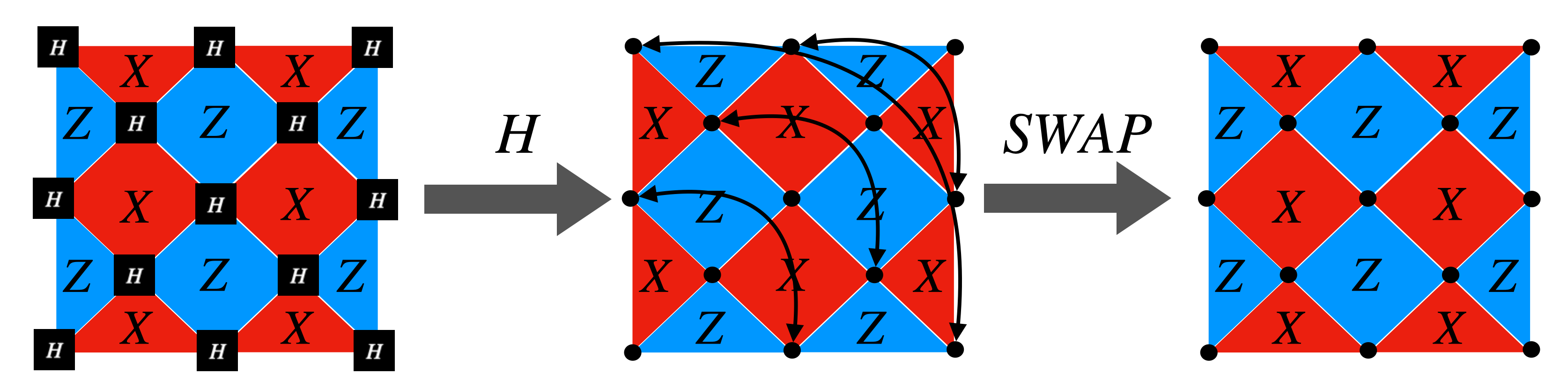}
    \caption{\textbf{Action of transversal $H$ on the surface-code stabilizers.}
    Action of the transversal $H_{XY}$, split between $Z$ (top) and $X$ (bottom) stabilizers. Each panel shows the state of the stabilizer group before the operation depicted in it, and the last panel shows the final form of the stabilizers. Double lines connect stabilizers with spatially separated support.}
    \label{fig:H}
\end{figure*}

For the $H$ operator, under conjugation, we have:
\[
X\mapsto Z,\quad  
 Z \mapsto X.
\]
To achieve this on the surface code, we the $H$ operator on all the physical qubits, and then reflect the code across its diagonal using $\text{SWAP}$ gates. As can be seen in Figure \ref{fig:H}, the effect of the $H$ gates is to exchanges $X$ and $Z$ stabilizers, while the $\text{SWAP}$ recovers the original structure of the surface code. Note that for every pair of dual stabilizer $S_X$ and $S_Z$, the result of the transformation is $S_X\mapsto S_Z$ and $S_Z\mapsto S_X$, which is an automorphism on the stabilizer group. The $X$ and $Z$ logical operators map to each other.

\subsection{Logical Scaling of CX Cultivation}
\label{SI:cx_cultivation}

The logical error rate of CX measurement has different properties compared with $H/H_{XY}$ measurement. In the $H/H_{XY}$ measurement schemes, the magic state is injected into the code, making any following measurement deterministic (absent any errors). 

Generally, injection improves the rate by reducing the probability of logical error by an additional order of $p$. It is similarly possible to inject a $\left|\text{CX}\right>$ state at the beginning of the cultivation procedure, but some errors at that stage would not be identifiable later. For example, a failed injection that resulted in the state $\left|0, +\right>$ would not be detected by any subsequent $CX$ measurement, since $\left|0, +\right>$ is a $+1$
 eigenstate of $CX$ but is not the $\ket{CX}$ state.
 
Instead, for $CX$ cultivation, the two surface codes are initialized in the $\ket{+,0}$ state by initializing each physical qubit in the $\ket{+}$ and $\ket{0}$ states, respectively, and measuring all stabilizers (see Appendix~\ref{SI:injection}). The $CX$ measurement is repeated $d_1$ times, and the state is discarded following any detection event. The initial measurement projects the state to the $\ket{\text{CX}}$ state with probability 3/4, and any subsequent measurement is deterministic (again, absent any errors). In this way, the $CX$ scheme has a fault distance $d_1$, at the expense of an additional $CX$ measurement and non-deterministic initialization. 

For the same reason, the codes cannot grow during the $CX$ cultivation stage. The $CX$ measurement projects into an eigenspac that contains multiple states, so some weight $3$ errors aren't identifiable in the $d=5$ state. For example, consider a logical $Z$ error on the control qubit during the $d_1=3$ stage. This error has weight 3, but it will not be identifiable by a subsequent $d_1=5$ $CX$ measurement, as it commutes with CX. To achieve fault distance of $5$, the surface codes should be initialized and measured at $d_1=5$ for the entire cultivation protocol.

In Clifford simulation of $CX$ cultivation, our circuit measures the $XX$ operator on two surface codes instead of the CX operator. Just like the $CX$ projection, the $XX$ measurement projects a pair of codes into a subspace with an additional degree of freedom, namely the $ZZ$ operator, and not to a specific state like $H$ and $H_{XY}$ cultivation schemes. 

 Initialization in the Clifford simulation is achieved by initilizing the two surface code in the state $\left|+,0\right>$ state. We then apply a transversal $CX$ gate between them, resulting in the state:
 $$
 \left|00\right>+\left|11\right>,
 $$
 which is a $+1$ eigenstate of both of their logical $\text{X}_1\text{X}_2$ and $\text{Z}_1\text{Z}_2$ operators. We then apply a logical $Z_1$ operator with probability $1/4$, which simulates the probabilistic nature of the initial $CX$ projection. 

\begin{figure*}[t] 
    \centering

    \includegraphics[width=0.65\textwidth,trim = {0 800 0 0}]{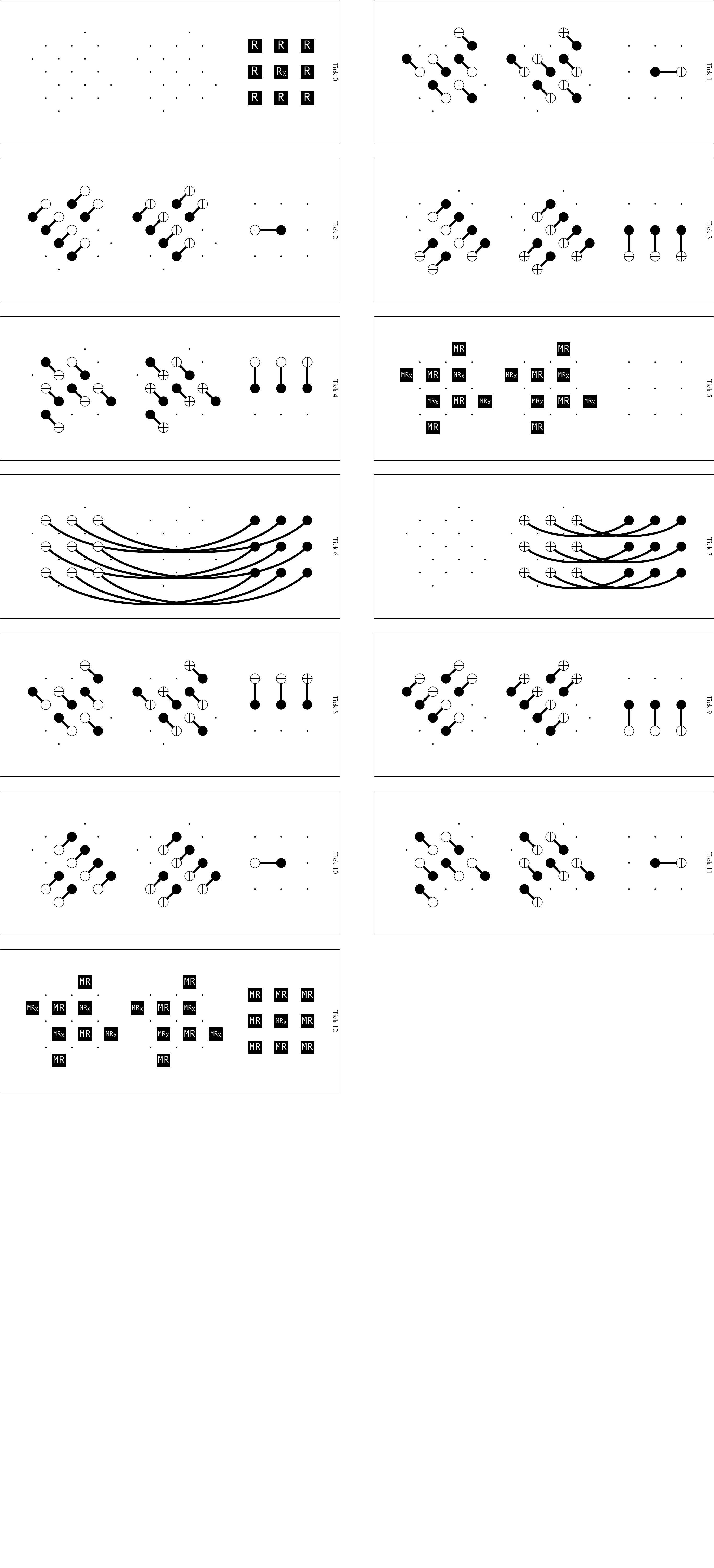} 
    \caption{\textbf{Circuit measuring the $XX$ operator} --
    \label{fig:transversal_cx_measurement_diagram}
    A GHZ measurement of the $XX$ operator on a pair of surface codes. From top to bottom, left to right, each panel represent a time slice of the circuit (labeled with tick index). During the collapse and expansion stages of the GHZ state, the two surface codes are performing their respective stabilizer measurements. To measure $CX$, Ticks 6-7 should show a transversal $CCX$ gates instead of $CXX$.}
    \href{https://algassert.com/crumble#circuit=Q(-0.5,0.5)0;Q(0,0)1;Q(0,1)2;Q(0,2)3;Q(0.5,0.5)4;Q(0.5,1.5)5;Q(0.5,2.5)6;Q(1,0)7;Q(1,1)8;Q(1,2)9;Q(1.5,-0.5)10;Q(1.5,0.5)11;Q(1.5,1.5)12;Q(2,0)13;Q(2,1)14;Q(2,2)15;Q(2.5,1.5)16;Q(3.5,0.5)17;Q(4,0)18;Q(4,1)19;Q(4,2)20;Q(4.5,0.5)21;Q(4.5,1.5)22;Q(4.5,2.5)23;Q(5,0)24;Q(5,1)25;Q(5,2)26;Q(5.5,-0.5)27;Q(5.5,0.5)28;Q(5.5,1.5)29;Q(6,0)30;Q(6,1)31;Q(6,2)32;Q(6.5,1.5)33;Q(8,0)34;Q(8,1)35;Q(8,2)36;Q(9,0)37;Q(9,1)38;Q(9,2)39;Q(10,0)40;Q(10,1)41;Q(10,2)42;R_34_35_36_37_39_40_41_42;RX_38;TICK;CX_0_2_5_9_11_14_17_19_22_26_28_31_13_10_8_4_15_12_30_27_25_21_32_29_38_41;TICK;CX_0_1_5_8_11_13_17_18_22_25_28_30_7_10_2_4_9_12_24_27_19_21_26_29_38_35;TICK;CX_5_3_11_8_16_15_22_20_28_25_33_32_9_6_7_4_14_12_26_23_24_21_31_29_41_42_35_36_38_39;TICK;CX_5_2_11_7_16_14_22_19_28_24_33_31_3_6_1_4_8_12_20_23_18_21_25_29_41_40_35_34_38_37;TICK;MR_6_10_4_12_23_27_21_29;MRX_0_5_11_16_17_22_28_33;TICK;CX_34_1_35_2_36_3_37_7_38_8_39_9_40_13_41_14_42_15;TICK;CX_34_18_35_19_36_20_37_24_38_25_39_26_40_30_41_31_42_32;TICK;CX_0_2_5_9_11_14_17_19_22_26_28_31_13_10_8_4_15_12_30_27_25_21_32_29_41_40_35_34_38_37;TICK;CX_0_1_5_8_11_13_17_18_22_25_28_30_7_10_2_4_9_12_24_27_19_21_26_29_41_42_35_36_38_39;TICK;CX_5_3_11_8_16_15_22_20_28_25_33_32_9_6_7_4_14_12_26_23_24_21_31_29_38_35;TICK;CX_5_2_11_7_16_14_22_19_28_24_33_31_3_6_1_4_8_12_20_23_18_21_25_29_38_41;TICK;MR_6_10_4_12_23_27_21_29_34_35_36_37_39_40_41_42;MRX_0_5_11_16_17_22_28_33_38}{Click here to view in Crumble}
\end{figure*}

Fig.  \ref{fig:transversal_cx_measurement_diagram} shows our phase-kickback implementation for the Clifford simulation, where the $CX$ cultivation with $CCX$ is replaced with $CXX$.

Laslty, we turn to our definition of our complementary gap in the CX Clifford simulation. At the end of that simulation, both $XX$ and $ZZ$ are  measured perfectly. The log-likelihood of the matching for all four possible values of $XX$ and $ZZ$ is evaluated, and the most likely option is chosen. The complementary gap is defined as the log-difference between the two most likely options.  The $ZZ$ operator is not measured during the protocol, but its initialization
is fault-tolerant. This simulates the additional degrees of
freedom of the $+1$ subspace of $CX$ that are not probed by the projections, but are initialized fault-tolerantly.

\section{Noise model}
\label{SI:Noise Model}
We define the noise model similarly to uniform noise model as defined in \cite{gidney2024magic}, adding an additional term describing the noise coming from 3-qubit operation. We introduce a 3-qubit depolarizing channel that applies one of the 63 (non-trivial) Pauli channels to the 3-qubits with equal probability $p/63$. The noise channels are defined in Fig. \ref{fig:noisechannels} and the uniform noise model, used throughout the paper, is defined in Fig. \ref{fig:uniform_noise_model}. This noise model is motivated by our decomposition of $CSWAP$ into $CCZ$, $CX$ and single qubit rotations, see Fig. \ref{fig:gate_decomposition_CSWAP}.

For Fig. \ref{fig:erasure-gap-decoding}, we used a simpler noise model, where single and two qubit depolarizing errors follow every single and two-qubit gate with the same probability $p$, with no idling or measurement errors.

\section{Alternative Noise Model for Rydberg Atoms}
\label{SI:rydbergatoms_noise}

Since our simulation assumes all-to-all connectivity, it would be most efficient to implement on a device with long range gates such as a Rydberg Atoms based quantum computer \cite{saffman2010quantum, bluvstein2024logical}.  Qubits based on Rydberg atoms can have coherence time much longer than their gate time. The dephasing error rate is dominated by decacy during the excitation to the Rydberg state, which only happens during 2-Qubit gates. To label this effect, we changed the noise model of Fig. \ref{fig:uniform_noise_model} by removing any idling error, and reducing the probability of single qubit errors to $p/10$. Since our scheme rarely uses 1-qubit gates to begin with, the dominant effect is removing the idling errors. 

In Fig. \ref{fig:neutral_atoms_noise_Model}, we can see improved logical fidelities, as high-weight errors are significantly less probable. As was demonstrated in \cite{levine2019parallel, evered2023high}, there is a native implementation of a CCX gate Rydberg atom. It is therefore reasonable to expect future experiments would reduce the overhead of implementing CCX compared with the standard combination of multiple single and two qubit gates.

\begin{figure*}
    \centering
    \begin{tabular}{|r|l|}
    \hline
    Noise channel & Probability distribution of effects
    \\
    \hline
    $\text{MERR}(p)$ & $\begin{aligned}
        1-p &\rightarrow \text{(report previous measurement correctly)}
        \\
        p &\rightarrow \text{(report previous measurement incorrectly; flip its result)}
    \end{aligned}$
    \\
    \hline
    $\text{XERR}(p)$ & $\begin{aligned}
        1-p &\rightarrow I
        \\
        p &\rightarrow X
    \end{aligned}$
    \\
    \hline
    $\text{ZERR}(p)$ & $\begin{aligned}
        1-p &\rightarrow I
        \\
        p &\rightarrow Z
    \end{aligned}$
    \\
    \hline
    $\text{DEP1}(p)$ & $\begin{aligned}
        1-p &\rightarrow I
        \\
        p/3 &\rightarrow X
        \\
        p/3 &\rightarrow Y
        \\
        p/3 &\rightarrow Z
    \end{aligned}$
    \\
    \hline
    $\text{DEP2}(p)$ & $\begin{aligned}
        1-p &\rightarrow I
        \\
        p/15&\rightarrow\left\{ A\otimes B\mid A,B\in\left\{ I,X,Y,Z\right\} ,A\otimes B\neq I\otimes I\right\}
    \end{aligned}$\\
    \hline
    $\text{DEP3}(p)$ & $\begin{aligned}
        1-p &\rightarrow I
        \\
p/63&\rightarrow\left\{ A\otimes B\otimes C\mid A,B,C\in\left\{ I,X,Y,Z\right\} ,A\otimes B\otimes C\neq I\otimes I\otimes I\right\}    \end{aligned}$\\
    \hline
    \end{tabular}
\caption{\textbf{Definition of the various noise channels in the Clifford simulation} -- \label{fig:noisechannels} The table was adapted from \cite{gidney2024magic} with the addition of DEP3. }

\end{figure*}

\begin{figure*}
    
    \begin{tabular}{|r|l|}
    \hline
    Ideal gate & Noisy gate
    \\
    \hline
    (single qubit unitary, including idle) $U_1$ & $\text{DEP1}(p) \cdot U_1$
    \\
    $CX$ & $\text{DEP2}(p) \cdot CX$
    \\
    $CH_{\text{XY}}$ & $\text{DEP2}(p) \cdot CH_{\text{XY}}$
    \\
    $C\bar{H}_{\text{XY}}$ & $\text{DEP2}(p) \cdot C\bar{H}_{\text{XY}}$
    \\
    $CCX$ & $\text{DEP3}(3p)\cdot\text{DEP2}_{12}(p)\cdot\text{DEP2}_{13}(p) \cdot CXX$\\
    $CSWAP\cdot CH^{\otimes2}$ & $\text{DEP3}(3p)\cdot\text{DEP2}_{12}(p)\cdot\text{DEP2}_{13}(p) \cdot CSWAP\circ CH^{\otimes2}$\\
    $CCZ$ & $\text{DEP3}(3p)\cdot\text{DEP2}_{12}(p)\cdot\text{DEP2}_{13}(p) \cdot CCZ$\\

    \hline
    (reset) $R_X$ & $\text{ZERR}(p) \cdot R_X$
    \\
    $R_Z$ & $\text{XERR}(p) \cdot R_Z$
    \\
    $M_X$ & $\text{DEP1}(p) \cdot \text{MERR}(p) \cdot M_X$
    \\
    $M_Z$ & $\text{DEP1}(p) \cdot \text{MERR}(p) \cdot M_Z$
    \\
    \hline
    \end{tabular}
    
    \caption{ \textbf{Uniform noise model} \label{fig:uniform_noise_model}  Each perfect operation is replaced by a noisy version, with the noise channels defined in Fig. \ref{fig:noisechannels}. The non-Clifford operations are replaced by $CX$'s which effectively measures the logical $X$ or $XX$ operators. The noise model doesn't change when replacing non-Clifford gates with Clifford gates.
    Subscripts in label the subset of qubits depolarized, with labeled as 1-Control, 2,3-Targets. Table adapted from \cite{gidney2024magic} with some additions.}
    
\end{figure*}

\section{Full vector sampling}
\label{SI:full_vector_simulation}

\begin{figure} 
    \centering
    \includegraphics[width=0.5\textwidth]{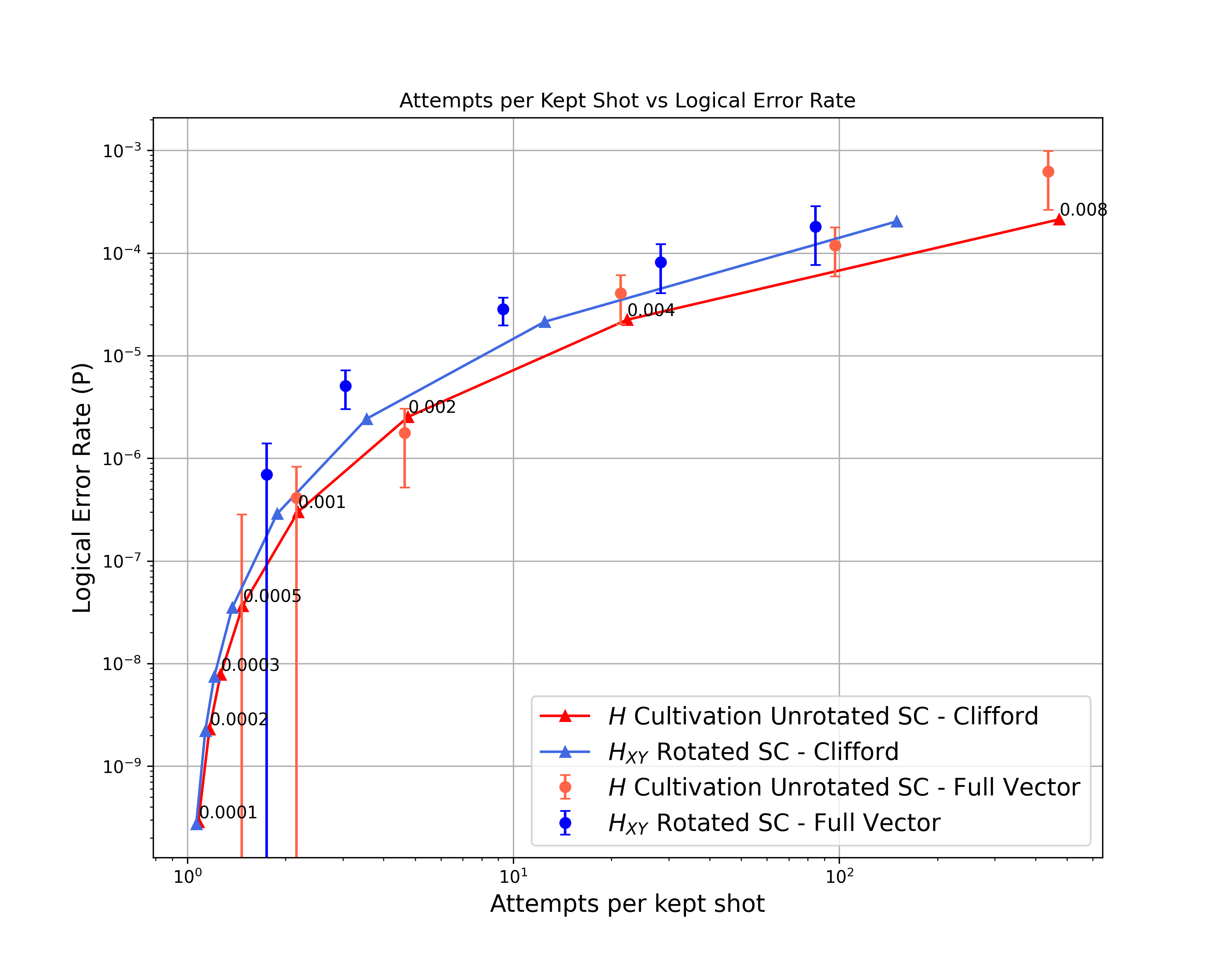}  
    
    \caption{{\textbf{Logical error rate of cultivation - comparing Clifford and Full Vector simulation} -- }
    \label{fig:full_vector_simulation}
    Logical error rate and acceptance rate of unrotated surface code $\text{H}$ cultivation with Uniform noise model using both a full vector simulation (black) and an approximation using a Clifford simulator. The simulation doesn't include the expansion step and post selects on all stabilizer measurements before a final (perfect) logical measurement in the Magic state basis.}
\end{figure}

We used a full vector sampling in order to benchmark our protocol spending roughly $10^6$ CPU hours resulting in roughly $10^7$ shots for each point. We only benchmarked the $H$ cultivation protocol using the unrotated surface code, and $H_{XY}$ protocol using the rotated surface code. The results are shown in Fig \ref{fig:full_vector_simulation}, in good agreement with the Clifford approximation, up to roughly a factor of 2.

We estimate the error bars in Fig \ref{fig:full_vector_simulation} by considering variance of our Bernoulli variable, which is $\text{Var}(X)=(1-p)p\approx p$. For $N$ accepted shots with $n_{error}$ errors, we set the error bars on our estimated $p$ to be:
\[
\sigma^2= \frac{n_{errors}}{N^2}
\]

\section{Preparation and Measurement of the GHZ Ancilla state}
\label{SI:GHZ}
 The GHZ ancilla state is generated by first initializing all of its qubits in the $\ket{0}$ state beside a single \textit{middle} qubit in the $\ket{+}$ state, and expanding the state in stages by repeated application of $CX$ as seen in Fig. \ref{fig:cultivation_digram} and in \cite{chamberland2020very}. 

 As mentioned in the main text, the measurement is just the expansion step done in reverse. This is equivalent to the measurement in \cite{chamberland2020very}, but requires a few additional time steps. In \cite{chamberland2020very}, the stabilizers of the repetition code are measured in separate flag qubits, followed by measurement of all the qubits composing the GHZ state in X basis. Here, the application of the expansion step means that the stabilizer information can be measured by measuring the non-\textit{middle} qubit in the Z basis. The logical measurement of the GHZ state can be recovered by measuring the \textit{middle} qubit in the X basis.

 Because of this equivalence, our GHZ measurement behaves as a one-flag circuit as defined in \cite{chamberland2020very}.

\section{Cultivation with Erasure Qubits}
\label{SI:erasure}
The benefit of using erasure qubits for cultivation is estimated analytically by assuming perfect detection and estimating the resulting acceptance rate. Any missed detection can be regarded as an additional Pauli error simply depolarizing the erased gate without a signal, which we assume for measurement error $q$, erasure rate $e$ and Pauli rate $p$ obeys $eq \ll p$. To simulate erasure, we define the Pauli noise level by a parameter $p$, as defined in Appendix \ref{SI:Noise Model}. At every noise location, we introduce additional erasure error with probability $e$. We assume the state is post-selected out following every erasure error. 

Fig. \ref{fig:erasure} shows the acceptance rate for cultivation on both the surface code for various values of $p,e$, in a system with erasure bias. To compare the benefit of erasure qubits, we assume there is some additional cost associated with erasure detection \cite{gu2024optimizing}. A non-erasure qubit circuit with Pauli error rate $p_0$, is compared to an erasure circuit with erasure $e$ and Pauli rate $p$ with more overall noise ($p+e>p_0$) and  bias for erasure ($p\ll p_0$). For a non-erasure qubit with Pauli rate $p_0=10^{-3}$, we can compare to $e\in \{1\times 10^{-3}, 2\times 10^{-3}\}$ erasures and $p=10^{-4}$ residual Pauli noise. For $e=2 p_0$, we say that the "overhead'' due to erasure conversion is two, and we see in Fig. \ref{fig:erasure} that the effect on $d_1=3,$ $H_{XY}$ surface code cultivation is to reduce the rate by a factor of $5$. At this cost, the logical fidelity improves by $\sim(p/p_0)^{d_1}\approx 10^3$. The acceptance rate decays exponentially in both qubit count and erasure rate. Therefore, doubling the erasure rate squares the required shots. 

We found the expansion stage very difficult to probe numerically. Introducing erasure into circuits for decoding requires generating a new matching graph for each circuit, which is costly. It is possible to probe higher noise rates which increase the logical error rates, and would then require fewer shots to estimate, but that too required too many shots due to the increased post-selection rate. 

A coarse but informative estimate of the post-selection overhead associated with the expansion step can be obtained by comparing the relative overhead between the initialization and final gap-estimation stages for the case $p = e$. This approach assumes that erasure errors have, on average, a comparable effect on gap decoding, while allowing for a much simpler numerical evaluation. When combined with the initialization acceptance rates presented in Fig.~\ref{fig:erasure}, this approximation provides a lower bound on the overall cultivation rate under erasure-biased noise. It is reasonable to assume erasure errors result in higher overall logical fidelity at the expense of less post-selection on the complementary gap. Erasure errors convey more information that can be utilized by the decoder (which is why the erasure threshold is much higher than the Pauli threshold for circuit model noise \cite{kubica2023erasure}). 

In order to demonstrate this effect, we used a simpler model that does not include post-selection on the stabilizers. A standard surface-code memory experiment is post-selected using only the complementary gap following $d$ rounds, with various values of erasure probability $e$ and Pauli noise $p$ in the same manner described earlier. The noise model is defined in Appendix \ref{SI:Noise Model} . As can be seen in Fig. \ref{fig:erasure-gap-decoding},  even when $e=2p$, the gap estimation can identify a correct decoding much more easily. 

\begin{figure}
\centering
        \includegraphics[totalheight=0.34 \textwidth]{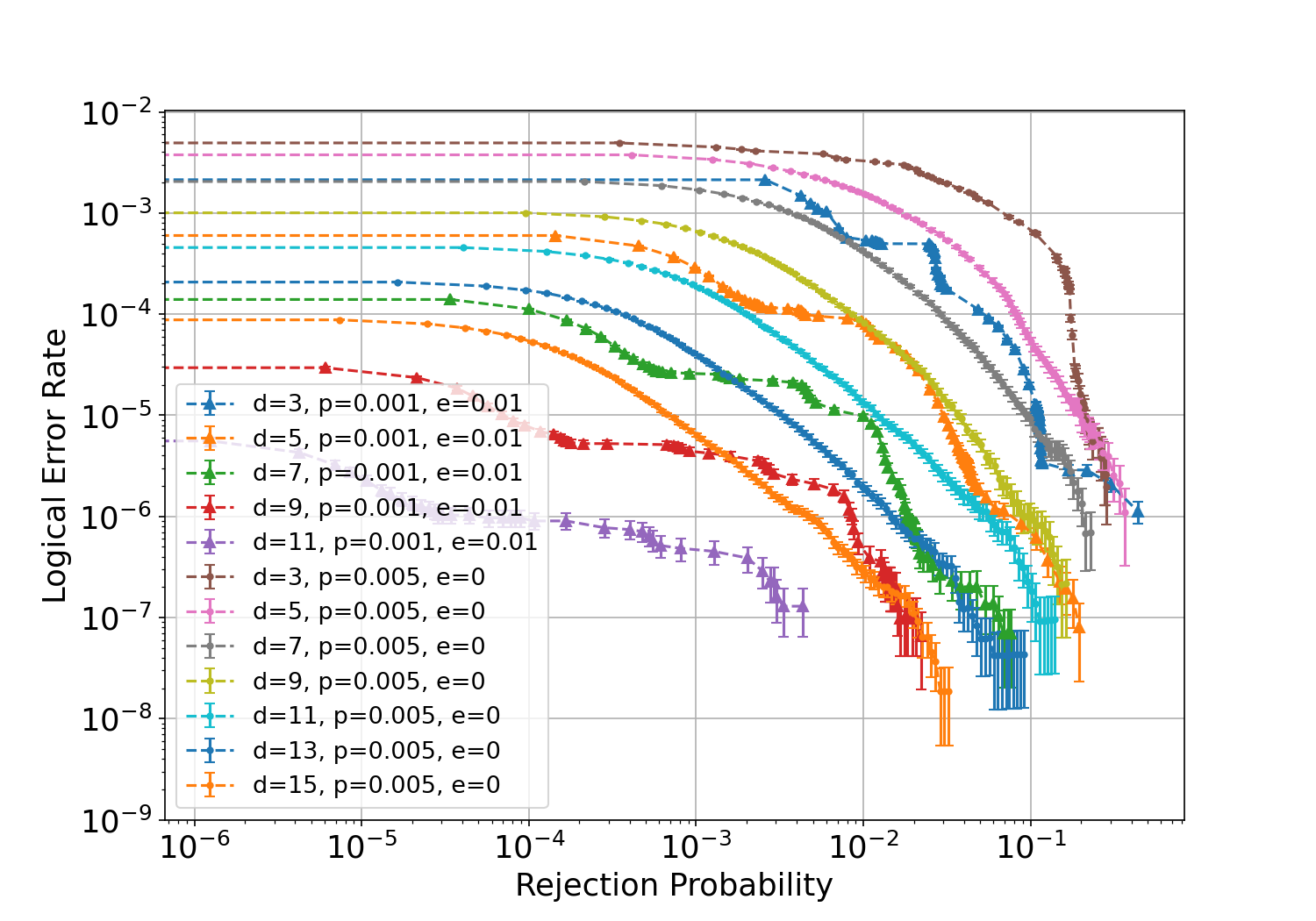}
    \caption{\textbf{Logical error rate and rejection probability of surface code memory experiment} --
    \label{fig:erasure-gap-decoding}
Logical error rate of $d$ rounds memory experiments. Different points label different threshold for the complementary gap. We compare two different regimes, one with only Pauli noise and one with mix of erasure and Pauli noise obeying $e=2p$. When erasure dominates the noise model, it can achieve the same logical fidelity and significantly higher rate with a factor of $2$ overhead.} 
\end{figure}

\section{Cultivation with Erasure qubits and $d_1=2$}
\label{SI:small_erasure}

\begin{figure} 
    \centering
    \includegraphics[width=0.5\textwidth]{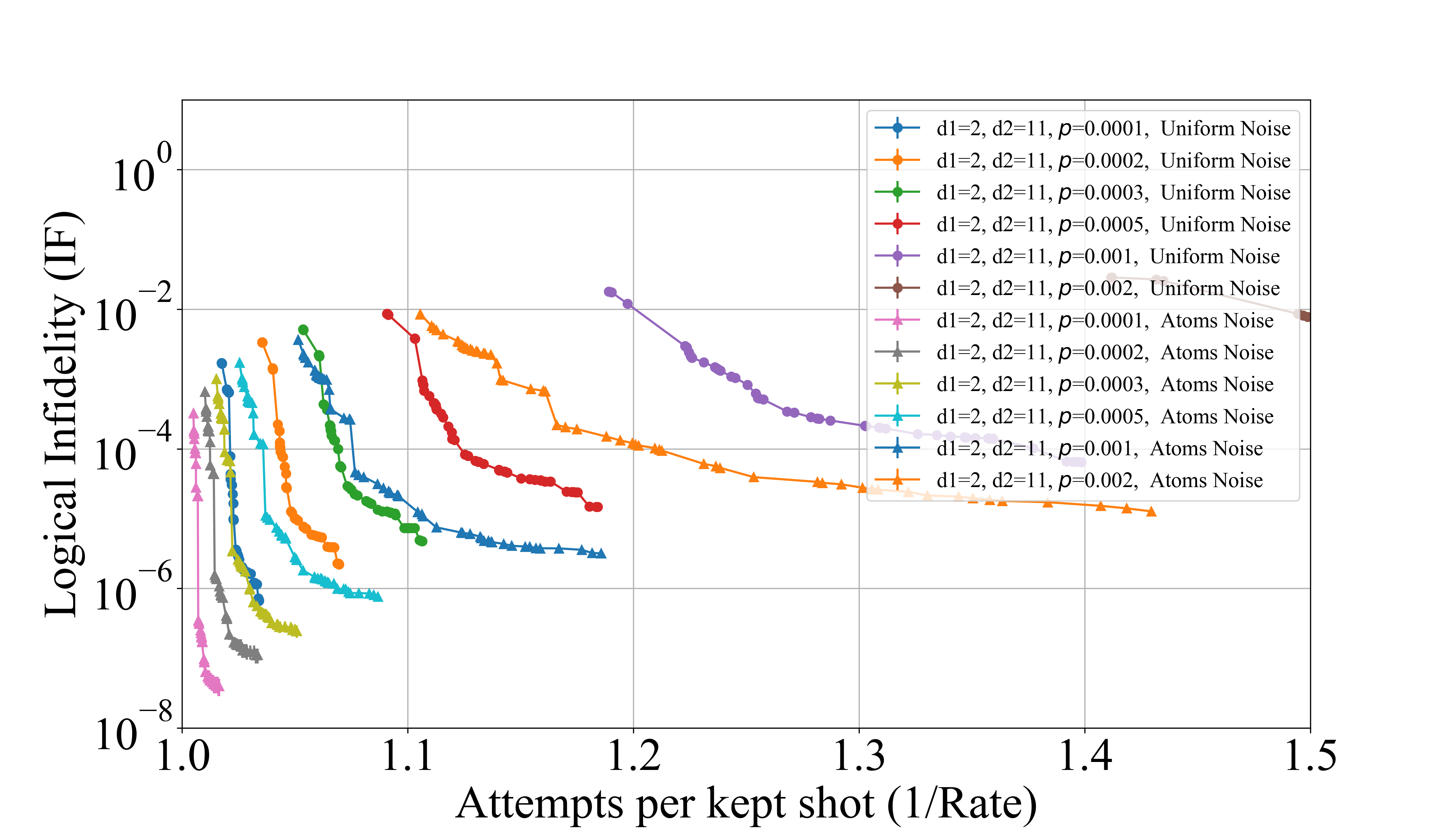}  
    
    \caption{{\textbf{Logical error rate of $H$ cultivation with expansion for $d_1=2$} -- }
    \label{fig:erasure_d2}
    Logical error rate and acceptance rate of unrotated surface code $\text{H}$ cultivation with Uniform and Atom noise model for $d_1=2$, following an expansion to $d_2=11$ unrotated surface code.}
\end{figure}

With erasure qubits, the residual Pauli error can reach low error rates. This suggests a simple proof of concept experiment for small devices with erasure qubits. Using a $d_1=2$ unrotated surface code that contains $5$ data qubits and $4$ ancilla, the logical $H$ operator can be measured using a single 3-qubit gate (with the ancilla's doubling as GHZ ancilla). Because of the small distance of the code, only a single GHZ measurement is required. We simulated this protocol without erasure noise for various values of the Pauli noise using both of our noise models. The simulation includes the expansion to $d_2=11$. If we want to estimate how such experiment will behave with $e=10^{-3}$ and $p=10^{-4}$, the logical fidelity will approximately be the same as our simulation with $p=10^{-4}$, and the rate can be deduced from the $p=10^{-3}$ simulation, see Fig. \ref{fig:erasure_d2}. With these assumptions, we expect to reach $10^{-6}$ infidelity with rate exceeding $70\%$ of shots. Without idling noise, we can further reach infidelities of $10^{-7}$ with over $85\%$ acceptance rate.

\section{CX cultivation with codes that encode multiple qubits}

Since $CX$ is transversal for all CSS codes, it can be used on pairs of codes that encode more than a single logical qubit. The eigenvalues of the transversal $CX$ operator would still be $\pm 1$, but measuring it would not project every single logical qubit into the $\ket{CX}$ state. Instead, a $+1$ measurement would project the qubits into an entangled state. 

For a pair of $[[n,k,d]]$ quantum codes,  the projection operator $P_{\pm}^i$ projects two logical qubits, one from each code, to the $\pm 1$ eigenspace of their combined logical $CX$ operator. The index $i$ labels the index of \textit{both} qubits, each at their respective codes. If all of the stabilizers of both codes are initialized in the $+1$ eigenstate, then a $+1$ measurement of the transversal $CX$ operator would effectively apply the following projection to the state:

\begin{equation}
\sum_{a\subseteq N}\prod_{i\notin a}P_{+}^{i}\prod_{j\in a}P_{-}^{j},
\end{equation}
where $N=\left\{ a\mid a\subseteq\{1...k\},\left|a\right|\text{ is even}\right\}$, i.e. all  sets of qubits with even size. This is different from projecting all pairs of qubits into the $\ket{\text{CX}}$ state, which is achieved by the $\prod_{i\le k}P_{+}^{i}$ operator. 

More explicitly, define the state $\ket{\overline{\text{CX}}}=\ket{1,-}$ state, which is the single eigenstate with eigenvalue $-1$ of the $CX$ operator. Projecting two $k=2$ codes into the $+1$ eigenstate of the transversal $CX$ operator would project the state into the following state:
\begin{equation}
\left|\text{CX}\right>\otimes\left|\text{CX}\right>+\left|\overline{\text{CX}}\right>\otimes\left|\overline{\text{CX}}\right>.
\end{equation}
Since this state isn't composed of two copies of the $\ket{\text{CX}}$ state, the method described in \cite{dennis2001toward} can't directly translate it into the CCX state. It is possible that future schemes will allow it to be used to generate multiple useful magic states, that can effectively increase the rate of the cultivation procedure. 

\section{3-Qubit gate compilation}
\label{3qubit_gate_compilation}
Our cultivation schemes make use of three types of three-qubit gates. As noted above, certain trapped-atom platforms natively support $CCZ$ gates, which simplify the implementation of $CCX$. For the $CSWAP$ gate, we employed the compilation from \cite{gouzien2025provably} (see Fig.~\ref{fig:gate_decomposition_CSWAP}), yielding an optimal decomposition that requires only two additional $CX$ gates.

\begin{figure}

    \includegraphics[width=0.5\textwidth]{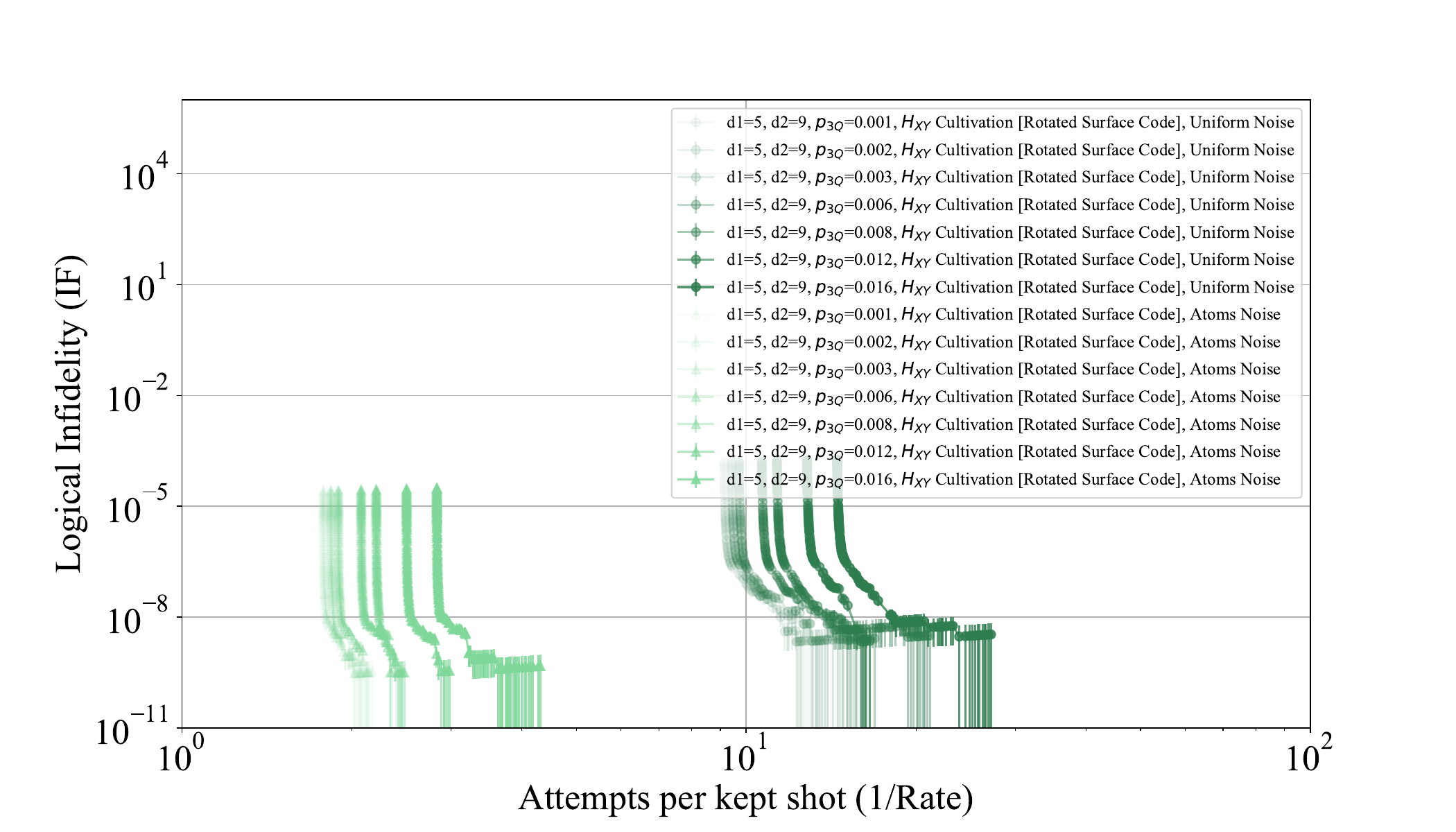}  
\label{fig:sensitivity_to_p3q}
\caption{\textbf{Logical fidelity and acceptance rate as a function of $p_{3Q}$.}}
    Logical error rate and acceptance rate of rotated surface code $H_{XY}$ cultivation with Uniform and Atom noise model, with varing strength for $p_{3Q}$ and constant $p=10^{-3}$. In the main text, all of our simulations used $p_{3Q}=3p=3\times10^{-3}$, as described in \ref{SI:Noise Model}.

\end{figure}

In our noise model, each three-qubit gate is followed by a $\text{DEPOLARIZE}3(3p)$ channel with probability $3p$, where $p$ denotes the two-qubit gate error rate for the rest of the circuit. Since three-qubit gates are used sparingly in the protocol, the overall performance is not highly sensitive to this assumption. Figure~\ref{fig:sensitivity_to_p3q} shows the effect of varying the noise strength of the $\text{DEPOLARIZE}3(p_{3Q})$ term while fixing $p=10^{-3}$. Even when $p_{3Q}$ is increased by more than an order of magnitude, the rate and fidelity are affected by less than a factor of two.
\begin{figure}
\[
\Qcircuit @C=1em @R=1em {
&\qw & \qw         & \qw     & \ctrl{2} & \qw       &\qw       & \qw & &   & & \ctrl{1}     & \qw \\
&\qw & \targ       & \qw     & \ctrl{1} & \qw       & \targ    & \qw & & = & & \qswap    & \qw \\
&\qw & \ctrl{-1}   & \gate{H}& \gate{Z} & \gate{H}  &\ctrl{-1} & \qw & &   & & \qswap \qwx & \qw \\
}
\]
    \caption{\textbf{Circuit Decomposition of 3-Qubit Controlled SWAP} in terms of Clifford and CCZ gates.}
    \label{fig:gate_decomposition_CSWAP}

    Circuit decomposition of Controlled SWAP gate using native gates for cold atoms - a single CCZ and two $CX$ gates. We verified that this decomposition is in fact optimal using the code provided in \cite{gouzien2025provably}.

\end{figure}

\section{Initialization and injection}
\label{SI:injection}

Here, we provide here additional details on the initialization steps of our cultivation protocols. Each protocol begins by resetting all data qubits to either the $\ket{0}$ or $\ket{+}$ state. Specifically, for $H$ cultivation we initialize in $\ket{0}$; for $H_{XY}$ we initialize in $\ket{+}$; and for CX cultivation we set the control patch to $\ket{+}$ and the target patch to $\ket{0}$.

As is standard for CSS codes, this initialization fixes half of the stabilizers to the $+1$ eigenstate—namely, those aligned with the basis of the initialized data qubits. The remaining stabilizers are left in random eigenstates. To enforce that all stabilizers are in the $+1$ eigenstate, we perform a syndrome extraction step to identify the randomly assigned stabilizers and then apply conditional single-qubit gates to flip any $-1$ outcomes to $+1$.

Following these corrections, we proceed with state injection. For $H_{XY}$ cultivation, we collapse the weight-3 logical $Z$ operator onto a single qubit using a pair of $CX$ gates. We then rotate that qubit by $\pi/8$ around the $Z$ axis and apply the $CX$ gates in reverse, preparing the surface-code patch in the $\ket{T}$ state.

For $H$ cultivation, we instead perform two successive rotations: a $\pi/8$ rotation about the logical $X$ axis, followed by a $\pi/4$ rotation about the logical $Z$ axis. These operations initialize the surface-code patch in the $\ket{T_H}$ state.

We then perform an additional syndrome extraction round, and post select on any faulty syndrome. 

\section{Insensitivity to $d_2$ and post-expansion rounds}
\label{SI:post_expansion_rounds}

After the final cultivation measurement, we expand each patch from the post-selected distance-$d_1$ code to a larger distance-$d_2$ code and then perform additional rounds of stabilizer measurements with soft decoding. In the main text we follow the convention of Ref.~\cite{gidney2024magic} and include 10 extra syndrome-extraction (SE) rounds after expansion as a simple model of decoder latency.

The role of the post-expansion rounds is to produce a full-distance syndrome history for the enlarged patch and to enable reliable soft-decoding of the final state via the complementary gap. For $d_2 \ge 2d_1+1$ the expanded code is large enough to support the subsequent fault-tolerant computation without changing the asymptotic logical scaling set by $d_1$.

Empirically, we find that the acceptance rate and the output fidelity are essentially insensitive to the precise value of $d_2$ and to the number of post-expansion rounds. The intuitive reason is that the dominant post-selection events originate from faults occurring on the original distance-$d_1$ region. 

\begin{figure}[t]
  \centering
  \includegraphics[width=\linewidth]{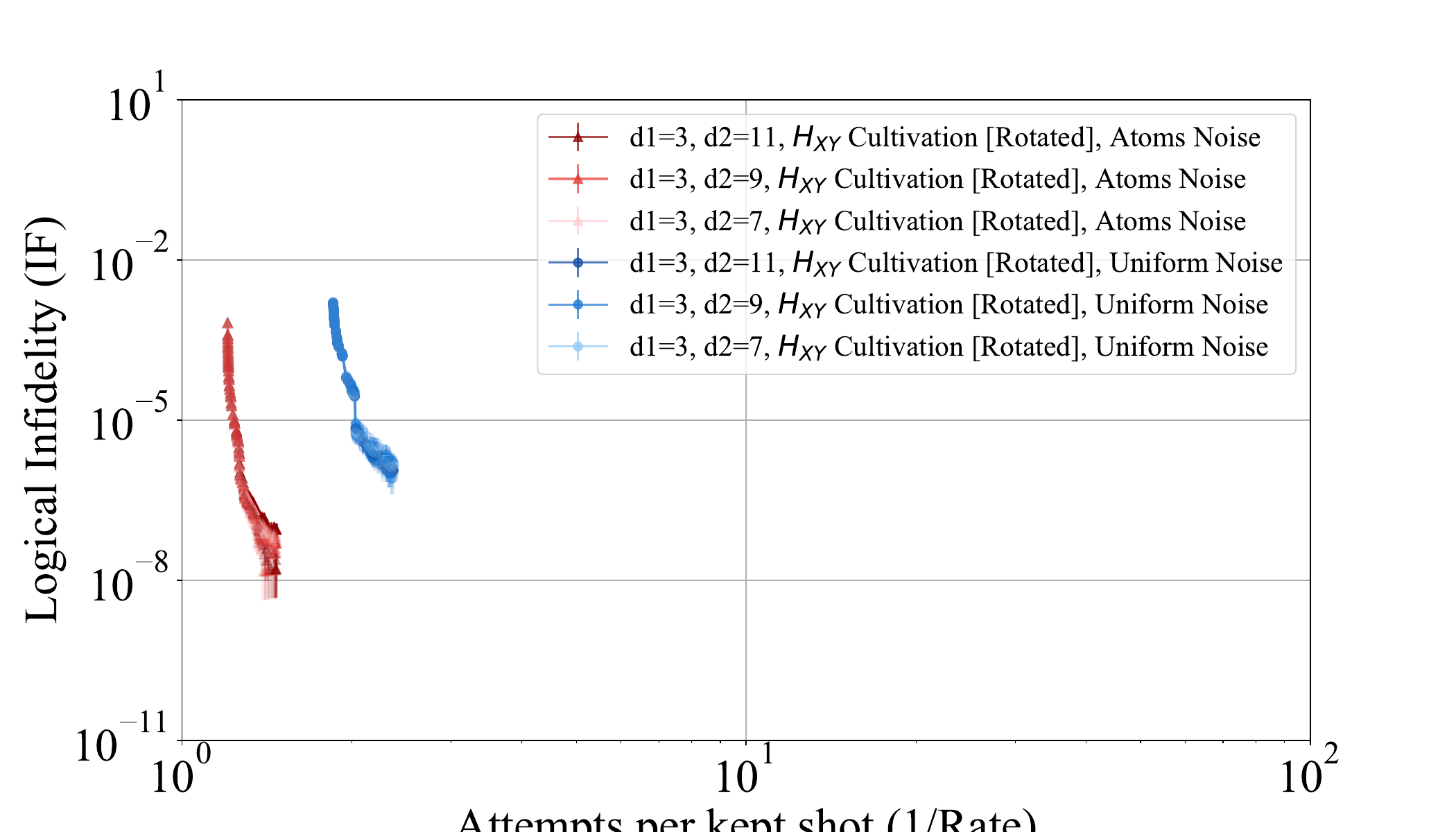}
  \caption{
  Sensitivity of the post-expansion stage to the choice of $d_2$ and to the number of post-expansion rounds $r$ for $d_1=3$ H$_{XY}$ cultivation on a rotated surface code.
  We plot the logical infidelity (IF) versus the number of attempts per kept shot ($1/\mathrm{Rate}$) while varying the expansion distance
  $d_2 \in \{7,9,11\}$, for both the atom and uniform noise models. For each $d_2$, we sweep the post-selection threshold on the decoder's
  complementary gap and repeat the sweep for multiple values of the number of post-expansion syndrome-extraction rounds $r$.
  All curves for different $r \in \{4, ... ,d_2\}$ are overlaid in the same panel using low opacity for readability.
  The resulting tradeoff curves are all within the variance of the sampling noise, across $d_2$ and $r$ with no observable trend. This  indicates  that once $d_2 \ge 2d_1+1$ and sufficient post-expansion syndrome information is collected, the acceptance rate and output fidelity are largely insensitive to the specific post-expansion parameters.
  }
  \label{fig:varying_d2_rounds}
\end{figure}

Figure~\ref{fig:varying_d2_rounds} illustrates this effect for $d_1=3$ H$_{XY}$ cultivation on a rotated surface code. We vary the expansion distance $d_2 \in \{7,9,11\}$ and also vary the number of post-expansion syndrome-extraction rounds, $r \in \{4, ..., d_2\}$. To present the full sweep compactly, we overlay all curves for different $r$ on the same axes with low opacity. The resulting tradeoff curves between logical infidelity and attempts per kept shot are nearly unchanged across both $d_2$ and $r$, for both the uniform and atom noise models. This supports treating acceptance rate and output fidelity as the most portable metrics across architectures, while qubits$\times$rounds remains strongly dependent on $d_2$ and $r$.

\section{Full protocol as psudo-code}
\label{SI:psudocode}
Below we provide pseudocode for the $H$ cultivation protocol on the unrotated surface code. In the rotated surface code, each double-check is performed by inserting it between the two halves of a syndrome extraction CX cycle. We follow this by post-selecting on the absence of any $-1$ syndrome outcomes. 
\hfill \break
\begin{algorithm}[H]
\caption{Pseudocode for Cultivation on an Unrotated Surface Code}
\begin{algorithmic}[1]
\Require Projection operator $P \in \{H, H_{XY}\}$, code distance parameter $d_1 \in \{3,5\}$, complementary gap threshold $\tau$, final distance $d_2$
\State Initialize a $d=3$ unrotated surface code
\If{$P = H$}
    \State Initialize data qubits in $\ket{0}$
\ElsIf{$P = H_{XY}$}
    \State Initialize data qubits in $\ket{+}$
\EndIf

\State Run a single round of syndrome extraction
\State Post-select on $Z$ stabilizers (for $H$) or $X$ stabilizers (for $H_{XY}$) in the $-1$ state

\If{$P = H$}
    \State Apply $Z$ corrections to set all $X$ stabilizers to $+1$
    \State Rotate around the logical $X$ axis by $\pi/8$
    \State Rotate around the logical $Z$ axis by $\pi/4$
\ElsIf{$P = H_{XY}$}
    \State Apply $X$ corrections to set all $Z$ stabilizers to $+1$
    \State Rotate around the logical $Z$ axis by $\pi/8$
\EndIf

\State Run a single round of syndrome extraction
\State Post-select on any stabilizers in the $-1$ state
\State \textit{Double-check} the logical operator 

\If{$d_1 = 5$}
    \State Expand the code to $d=5$ with all stabilizers in the $+1$ eigenstate using a unitary circuit
    \State Run a single round of syndrome extraction
    \State Post-select on any stabilizers in the $-1$ state
    \State \textit{Double-check} the logical operator
\EndIf

\State Expand the code to $d_2$ 
\State Run 10 rounds of syndrome extraction
\State Decode and post-select using the complementary gap threshold $\tau$
\end{algorithmic}
\end{algorithm}

\end{document}